\documentclass[prb,twocolumn,showpacs,floatfix]{revtex4}

\usepackage{graphicx}
\usepackage{amsmath}

\begin{document}

\title{Exchange Interaction Between Three and Four Coupled Quantum Dots:
Theory and Applications to Quantum Computing}
\author{Ari Mizel}
\affiliation{Physics Department and Materials Research Institute,
  Pennsylvania State University, University Park, PA 16802}
\author{Daniel A. Lidar}
\affiliation{Chemical Physics Theory Group, Chemistry Department, University of Toronto, 80
  St. George St., Toronto, Ontario M5S 3H6, Canada}

\begin{abstract}
Several prominent proposals have suggested that spins of localized electrons
could serve as quantum computer qubits. The exchange interaction has been
invoked as a means of implementing two qubit gates. In this paper, we
analyze the strength and form of the exchange interaction under relevant
conditions. We find that, when several spins are engaged in mutual
interactions, the quantitative strengths or even qualitative forms of the
interactions can change. It is shown that the changes can be dramatic within
a Heitler-London model. Hund-M\"{u}lliken calculations are also presented,
and support the qualititative conclusions from the Heitler-London model. The
effects need to be considered in spin-based quantum computer designs, either
as a source of gate error to be overcome or a new interaction to be
exploited.
\end{abstract}

\pacs{03.67.Lx,75.10.Jm}
\maketitle

\section{Introduction}

The exchange interaction between electrons has been studied since the early
days of quantum mechanics \cite{Heisenberg:28,Dirac:29,vanVleck:book} and
has been reviewed in some classic
references,\cite{Herring:62,Anderson:63,Anderson:63a} as well as
textbooks.\cite{March:85vol1} Recently, a promising proposal \cite{Loss:98} has emerged to
use the exchange interaction as a tunable qubit-qubit interaction in a
quantum computer, with the individual spins of electrons acting as qubits.
To satisfy the conditions for constructing a universal quantum computer, the
exchange interaction can either be supplemented with single-qubit operations,
\cite{Loss:98} or can be used by itself to construct a universal set of
gates, in which case one encodes a logical qubit into the state of several
spins.\cite{Bacon:99a,Kempe:00,DiVincenzo:00a,LidarWu:01} (This alternative
to the standard universality scheme has been termed \textquotedblleft
encoded universality\textquotedblright .\cite{Bacon:Sydney}) Motivated by
the proposal of Loss and DiVincenzo \cite{Loss:98}, there have been a number
of studies of the one-particle and two-particle behavior of electrons
localized on quantum dots within a quantum computer. \cite{Loss:98,Burkard:99,Burkard:00,Hu:99,Hu:01b,Levy:01a,Schliemann:01,Kavokin:01,Barrett:02}
Here, we expand on our work \cite{MizelLidar:PRL04} considering the
important situation of three or more coupled dots. We show how both
quantitatively and qualitatively new effects can appear. These effects
require consideration if one intends to make a quantum computer with more
than two spins.

The exchange interaction between two localized electrons arises as a result
of their spatial behavior, but it can be expressed as an effective spin-spin
interaction. In conditions of rotation symmetry (i.e. neglecting external
magnetic fields, spin-orbit coupling, etc.), a purely isotropic form of this
interaction arises, which is known as Heisenberg exchange: 
\begin{equation}
H_{\mathrm{ex}}=J\mathbf{S}_{A}\cdot \mathbf{S}_{B}.  \label{eq:Hex}
\end{equation}
Here, $\mathbf{S}=(S_{x},S_{y},S_{z})$ is a vector of spin-$\frac{1}{2}$
angular momentum operators, and $A,B$ are indices referring to the location
of each electron. (We take spin operators to be dimensionless in this paper
-- $\hbar $ is excluded from their definition.) This Hamiltonian has a
spin-singlet eigenstate and degenerate spin-triplet eigenstates. \cite
{Heisenberg:28} The quantity $J$ is the exchange coupling constant, given by
the energy splitting between the spin-singlet and spin-triplet states, \cite
{Heisenberg:28,Burkard:99,Hu:99} 
\begin{equation}
J=\epsilon _{t}-\epsilon _{s}.  \label{eq:J}
\end{equation}
To date, studies of the exchange interaction in quantum computation have
focused on the case of two quantum dots.\cite{Loss:98,Burkard:99,Burkard:00,Hu:99,Hu:01b,Levy:01a,Schliemann:01,Kavokin:01,Barrett:02}
Starting from the simplest case of two electrons in singly-occupied dots in
the lowest orbital state, systematic generalizations have been introduced
and their effect on the exchange interaction studied. In particular,
researchers have analyzed the effect of double occupation,\cite{Hu:99,Schliemann:01,Barrett:02} higher orbital states,\cite{Burkard:99,Hu:99} and many-electron dots.\cite{Hu:01b} An accurate
numerical study reporting singlet-triplet crossing via magnetic field
manipulation in a lateral double quantum dot can be found in Ref.~
\onlinecite{Yonnouleas:02}. Neglecting spin-orbit coupling, these studies
have found increasingly accurate expressions for $J$, while focusing on the
definition of Eq.~(\ref{eq:J}). In the presence of spin-orbit coupling both
rotation and inversion symmetry are broken, and anisotropic corrections to $
H_{\mathrm{ex}}$ arise.\cite{Dzyaloshinski:58,Moriya:60,Kavokin:01}

In this work, we undertake a study of the case of three or four electrons,
each in a quantum dot. Once the system involves more than two electrons,
simultaneous multi-partite exchanges can occur. For three coupled dots
containing three electrons, processes in which all three electrons exchange
contribute to a quantitative correction to the value of $J$. We show explicitly
that, for three identical dots arranged on the corners of an equilateral
triangle, the effective Hamiltonian can still be written using a Heisenberg
exchange interaction 
\begin{equation*}
H_{\mathrm{spin}}=K + J\left( \mathbf{S}_{A}\cdot \mathbf{S}_{B}+\mathbf{S}
_{B}\cdot \mathbf{S}_{C}+\mathbf{S}_{C}\cdot \mathbf{S}_{A}\right) ,
\end{equation*}
but $J$ is now found to be influenced by three-body exchange matrix
elements. For four coupled dots containing four electrons, the actual form
of the interaction Eq.~(\ref{eq:Hex}) changes due to four-body effects. For
identical dots arranged on the corners of a symmetric tetrahedron, the
interaction takes the form 
\begin{eqnarray*}
H_{\mathrm{spin}} &=& K + J\sum_{A\leq i<j\leq D}\mathbf{S}_{i}\cdot \mathbf{
S} _{j} + J^{\prime } [ (\mathbf{S}_{A}\cdot \mathbf{S}_{B})(\mathbf{S}
_{C}\cdot \mathbf{S}_{D})  \notag \\
&+& (\mathbf{S}_{A}\cdot \mathbf{S}_{C})(\mathbf{S} _{B}\cdot \mathbf{S}
_{D})+(\mathbf{S}_{A}\cdot \mathbf{S}_{D})(\mathbf{S} _{B}\cdot \mathbf{S}
_{C}) ]
\end{eqnarray*}
where throughout this paper notation like $A\leq i \leq D$ means that $i$
takes the letter values $A$ to $D$. According to our Heitler-London (HL)
calculations, the ratio $|J^{\prime}/J|$ can reach $15\%$ in in physically
relevant parameter regimes.

Four-body exchange terms have been discussed in other contexts -- for
example, in a perturbative treatment of the two and the three dimensional
half-filled Hubbard models four-body interactions were shown to suppress the
N\'{e}el temperature and the temperature of the para-ferromagnetic phase
transition \cite{Takahashi:77,MacDonald:88} (see Appendix~\ref{app:pert}).
Here, we present a non-perturbative derivation of these terms, starting from
a finite-dimensional Hamiltonian, and then highlight their significance for
quantum computation. Interaction Hamiltonian calculations like ours are of
significance in various quantum computation contexts including (i) the
encoded universality paradigm, where in the most efficient implementations
several exchange interactions are turned on simultaneously \cite{Bacon:99a,Kempe:00,DiVincenzo:00a,LidarWu:01,Bacon:Sydney} (quantitative
studies of parallel gate sequences \cite{DiVincenzo:00a} in particular may
require revisiting in light of our results, as well as the \textquotedblleft
supercoherent qubits\textquotedblright\ method for reducing decoherence,\cite{Bacon:01} where four and eight-spin interactions must be turned on
simultaneously in order to enact quantum logic gates between encoded
qubits); (ii) adiabatic quantum computing \cite{Farhi:01}, where the final
Hamiltonian for any non-trivial calculation inevitably includes simultaneous
interactions between multiple qubits; (iii) fault-tolerant quantum error
correction, where a higher degree of parallelism translates into a lower
threshold for fault-tolerant quantum computation operations; \cite{Shor:96,Gottesman:97a,Preskill:97a,Steane:99a,Lidar:00b}; (iv) the
\textquotedblleft one-way\textquotedblright\ quantum computer proposal,\cite{Raussendorf:01} where all nearest-neighbor interactions in a cluster of
coupled spins are turned on simultaneously in order to prepare many-spins
entangled state; (v) the search for physical systems with intrinsic,
topological fault tolerance, where systems with four-body interactions have
recently been identified as having the sought-after properties.\cite{Freedman:02}

We begin with a general description of a finite-dimensional effective spin
Hamiltonian in Section~\ref{general-H}. (This is compared to the standard,
perturbative derivation in Appendix~\ref{app:pert}.) Section~\ref{parameters}
shows how to compute the parameters in the effective spin Hamiltonian, with
detailed consideration of the two-electron, three-electron, and
four-electron cases. We introduce a specific model and calculate the
parameters quantitatively, for three and four electrons, in
Section~\ref{values}. Appendices~\ref{app:3elec} and \ref{app:4elec}
contain relevant technical details.

\section{Electron-spin-operator Hamiltonian}

\label{general-H}

In this section we present general arguments concerning the form of the
effective spin Hamiltonian, as it arises from $n$ localized electrons
interacting via the Coulomb force. We start with the familiar electronic
Hamiltonian 
\begin{eqnarray}
H &=& \sum_{i=1}^{n}\frac{1}{2m}\mathbf{p}_{i}^{2}+V(\mathbf{r}
_{i})+\sum_{i<j} \frac{e^{2}}{\kappa |\mathbf{r}_{i}-\mathbf{r}_{j}|}  \notag
\\
&\equiv& \sum_{i=1}^{n}h( \mathbf{r}_{i})+\sum_{i<j}w(\mathbf{r}_{i},\mathbf{
r}_{j}),  \label{spatialHamiltonian}
\end{eqnarray}
where the first term is the kinetic energy, the second is the confining
potential, and the third is the Coulomb interaction. The confining potential 
$V(\mathbf{r})$ contains $n$ energy minima, which give rise to the $n$ dots.
To understand the dynamics of $n$ electron-spin qubits in $n$ quantum dots,
it is desirable to eliminate the spatial degrees of freedom, leaving an
effective Hamiltonian composed of electron-spin operators only.

The first step in changing Hamiltonian (\ref{spatialHamiltonian}) to an
electron-spin-operator Hamiltonian is to fix a basis. We first consider the
case of two electrons in two dots, labeled $A$ and $B$. We do not allow for
double occupancy of a dot and consider only a single low-energy orbital per
dot labeled as $\phi _{A}(\mathbf{r})\equiv \left\langle \mathbf{r}
|A\right\rangle $ and $\phi _{B}(\mathbf{r})\equiv \left\langle \mathbf{r}
|B\right\rangle .$ Electrons $1$ and $2$ occupy these low energy orbitals.
Each electron can have spin-up or spin-down, hence each electron can
represent a qubit. A state with, for example, electron $1$ in orbital $B$
with spin-up and electron $2$ in orbital $A$ with spin-down is represented
as $\left\vert BA \right\rangle \left\vert \uparrow \downarrow \right\rangle$. Since electrons are fermions, this state needs to be antisymmetrized; the
full state of the two electrons takes the form of a Slater determinant 
\begin{eqnarray}
\left\vert \Psi (\downarrow \uparrow )\right\rangle = (\left\vert
AB\right\rangle \left\vert \downarrow \uparrow \right\rangle -\left\vert
BA\right\rangle \left\vert \uparrow \downarrow \right\rangle ) \propto
a_{A\downarrow }^{\dag }a_{B\uparrow }^{\dag } \left\vert \mathrm{vac}
\right\rangle .  \label{eq:downup}
\end{eqnarray}
Note that the order of spins in the state label $\Psi (\downarrow \uparrow )$
indicates that the electron in orbital $A$ has spin down and the electron in
orbital $B$ has spin up. In Eq.~(\ref{eq:downup}) we introduced
second-quantized notation (ignoring normalization), with $a_{A\downarrow
}^{\dag }$ creating an electron with spin down in orbital $A$ and $
a_{B\uparrow}^{\dag }$ creating an electron with spin up in orbital $B$. The
four states $\left\vert \Psi (s_{A},s_{B})\right\rangle $ form the
two-electron basis. The same procedure applies to three electrons in three
dots. There is again a single low-energy orbital per dot, labeled as $\phi
_{A}(\mathbf{r} )\equiv \left\langle \mathbf{r}|A\right\rangle $, $\phi _{B}(
\mathbf{r} )\equiv \left\langle \mathbf{r}|B\right\rangle $, and $\phi _{C}(
\mathbf{r} )\equiv \left\langle \mathbf{r}|C\right\rangle $ for dots $A$, $B$, and $C$ respectively. Electrons $1$, $2$, and $3$ occupy these low energy
orbitals. A state with, for example, electron $1$ in orbital $B$ with
spin-up, electron $2$ in orbital $A$ with spin-down, and electron $3$ in
orbital $C$ with spin-up is represented as $\left\vert BAC\right\rangle
\left\vert \uparrow \downarrow \uparrow \right\rangle .$ This state is then
antisymmetrized so that the full state of the three electrons is the Slater
determinant 
\begin{eqnarray}
\left\vert \Psi (\downarrow \uparrow \uparrow )\right\rangle &\equiv &
(\left\vert ABC\right\rangle \left\vert \downarrow \uparrow \uparrow
\right\rangle +\left\vert BCA\right\rangle \left\vert \uparrow \uparrow
\downarrow \right\rangle +\left\vert CAB\right\rangle \left\vert \uparrow
\downarrow \uparrow \right\rangle  \notag \\
&-& \left\vert BAC\right\rangle \left\vert \uparrow \downarrow \uparrow
\right\rangle -\left\vert ACB\right\rangle \left\vert \downarrow \uparrow
\uparrow \right\rangle -\left\vert CBA\right\rangle \left\vert \uparrow
\uparrow \downarrow \right\rangle )  \notag \\
&\propto &a_{A\downarrow }^{\dag }a_{B\uparrow }^{\dag }a_{C\uparrow }^{\dag
}\left\vert \mathrm{vac}\right\rangle .  \label{eq:downupup}
\end{eqnarray}
The order of spins in the state label $\Psi (\downarrow \uparrow \uparrow )$
indicates that the electron in orbital $A$ has spin down, the electron in
orbital $B$ has spin up, and the electron in orbital $C$ has spin up. The
eight states $\left\vert \Psi (s_{A},s_{B},s_{C})\right\rangle $ form the
three-electron basis.

The general case of $n$ electrons gives $2^{n}$ fully antisymmetrized basis
vectors of the form 
\begin{equation}
\left\vert \Psi (s_{A},s_{B},\dots,s_{Z} )\right\rangle = \sum_{P}\delta
_{P}P[\left\vert AB\dots \right\rangle \left\vert s_{A}s_{B} \dots \label{eq:statedef}
\right\rangle ],
\end{equation}
where the sum runs over all permutations $P$ of both orbitals and spins, and 
$\delta _{P}=1$ $(-1)$ if the permutation is even (odd). In this basis, the
Hamiltonian (\ref{spatialHamiltonian}) takes the form of a $2^{n}\times
2^{n} $ Hermitian matrix. Like any $2^n \times 2^n$ Hermitian matrix, the
Hamiltonian can be written as a sum 
\begin{equation}  \label{spinsum}
H_{\mathrm{spin}} = \sum _{i,j,\cdots = 0}^3 l_{i,j,\cdots} \;\;\;\; \sigma
_{i}(A)\otimes \sigma _{j}(B)\otimes \cdots
\end{equation}
of Hermitian spin matrices of the form {$\sigma _{i}(A)\otimes \sigma
_{j}(B)\otimes \cdots$} each multiplied by a real coefficient $
l_{i,j,\cdots} $. Here, $\sigma_{i}(p)$ denotes the Pauli matrix $\sigma
_{i} $ acting on the electron in dot $p $, with $i =0,1,2,3$ and with $
\sigma _{0} $ equal to the identity matrix. There are $n$ factors in the
tensor product $\sigma _{i}(A)\otimes \sigma _{j}(B)\otimes \cdots$, so that
it can be written as a $2^n \times 2^n$ matrix, and there are similarly $n$
subscripts on the coefficient $l_{i,j,\cdots}$. This decomposition (\ref{spinsum}) into spin matrices produces an effective electron-spin
Hamiltonian that conveniently describes the dynamics of $n$ qubits.

The procedure we have just described is framed within the Heiter-London
approximation.\cite{Heitler:27} The approximation consists in neglecting
excited states and has been criticized on the grounds that it does not
produce the correct asymptotic behavior in the limit of very large distances.
\cite{Herring:62} However, in the context of our system of interest, this
asymptotic limit is not a concern, and moreover, recent studies have
verified the utility of the approximation in the case of large (but not
infinite) inter-dot separation.\cite{Hu:99} We will thus proceed with the HL
approximation, which has the advantage of conceptual simplicity and physical
clarity. In the three electron case, we show that Hund-M\"ulliken (HM)
calculations, in which double occupation is permitted, support the
conclusions of our HL results.

Symmetry considerations fundamentally constrain the form of the
electron-spin Hamiltonian. The coordinate system used to define $\uparrow $
and $\downarrow $ is arbitrary if there is no spin-orbit coupling and no
external magnetic field. In this case, the effective spin operator
Hamiltonian has rotation, inversion, and exchange symmetry. The coefficients 
$l_{i,j,\cdots }$ in (\ref{spinsum}) are strongly constrained by this
symmetry. The Hamiltonian can only be a function of the total spin squared $
\mathbf{S}_{T}^{2}=(\mathbf{S}_{A}+\mathbf{S}_{B}+\dots )^{2}$, where {$
\mathbf{S}_{A} \equiv \frac{1}{2} (\sigma_1(A) \hat{x} +\sigma_2(A) \hat{y}
+\sigma_3(A) \hat{z}) \otimes \sigma_0(B) \otimes \sigma_0(C) \otimes \cdots$}, {$\mathbf{S}_{B} \equiv \sigma_0(A) \otimes \frac{1}{2} (\sigma_1(B) \hat{
x} +\sigma_2(B) \hat{y} +\sigma_3(B) \hat{z}) \otimes \sigma_0(C) \otimes
\cdots$}, etc. A scalar such as $\mathbf{S}_{A}\cdot (\mathbf{S}_{B}\times 
\mathbf{S}_{C})$ cannot appear in the Hamiltonian because of inversion
symmetry. We must have 
\begin{equation}
H_{\mathrm{spin}}=L_{0}+L_{1}\mathbf{S}_{T}^{2}+L_{2}(\mathbf{S}
_{T}^{2})^{2}+\dots  \label{spinHamiltonian}
\end{equation}
where $L_{0},L_{1},L_{2},\dots $ are real constants with dimensions of
energy. The constant $L_{0}$ is an energy shift. The term proportional to $
L_{1}$ gives rise to the familiar Heisenberg interaction. Here we see that
in principle higher order interactions may be present in the spin
Hamiltonian, starting with a fourth order term proportional to $L_{2}$. In
this highly symmetric situation, the eigenstates of the spin Hamiltonian are
clearly just eigenstates of $\mathbf{S}_{T}$.

\section{Computation of the Spin Hamiltonian Parameters}

\label{parameters}

To compute the values of $L_0 ,L_1 ,L_2 ,\dots $ we consider an eigenstate $
|\Psi \rangle $ of $\mathbf{S}_{T}^{2}$, with known eigenvalue $
S_{T}(S_{T}+1)$. If there are $n$ electrons in the system, we write $|\Psi
\rangle = |\Psi _{S_{T}}^{n}\rangle$. To proceed, one (i)\ computes the
expectation value of the effective spin Hamiltonian (\ref{spinHamiltonian})
in this state, (ii)\ computes the expectation value of the spatial
Hamiltonian (\ref{spatialHamiltonian}) in this state, and then (iii) equates
the two expectation values: 
\begin{equation}
\langle \Psi |H_{\mathrm{spin}}|\Psi \rangle =\langle \Psi |H|\Psi \rangle .
\label{eq:equal}
\end{equation}
This procedure is repeated for all eigenvalues of $\mathbf{S}_{T}^{2}$, thus
generating a set of linear equations for the parameters $L_0 ,L_1 ,L_2
,\dots $, in terms of matrix elements of $H$ between different orbital
states. For $n$ electrons the number of distinct eigenvalues of $\mathbf{S}
_{T}^{2}$ is $\lfloor \frac{n}{2} \rfloor+1 $ (where $\lfloor \frac{n}{2}
\rfloor$ denotes the greatest integer less than $\frac{n}{2})$, so this is
the maximum number of distinct energy eigenvalues of the Hamiltonian (\ref
{spinHamiltonian}). Thus, the coefficients $L_m$ for $0\le m < \lfloor \frac{
n}{2} \rfloor + 1$ have enough degrees of freedom to completely and uniquely
specify the matrix (\ref{spinHamiltonian}); without loss of generality, we
can set $L_m = 0$ for $m \ge \lfloor \frac{n}{2} \rfloor + 1$. We are led to 
$\lfloor \frac{n}{2} \rfloor+1$ coupled linear equations for the non-zero $
L_m$ parameters. In the case that $n$ is even $S_{T}$ takes on the integer
values $0,1,\dots ,n/2 $. In the case that $n$ is odd $S_{T}$ takes on the
half-integer values $1/2,3/2,\dots ,n/2$. Then 
\begin{eqnarray}
\langle \Psi _{S_{T}}^{n}|H_{\mathrm{spin}}|\Psi _{S_{T}}^{n}\rangle
&=&\sum_{m=0}^{\lfloor \frac{n}{2} \rfloor}L_{m}\left( S_{T}(S_{T}+1)\right)
^{m}  \label{eq:Hspin-even}
\end{eqnarray}
Having completed step (i) of our program, we now turn to step (ii), the
calculation of $\langle \Psi _{S_{T}}^{n}|H|\Psi _{S_{T}}^{n}\rangle $. We
make this calculation separately for the cases of two, three and four
electrons.

\subsection{Two Electron Case}

As a simple illustration of our procedure we rederive the well-known result
for two electrons: the exchange constant equals the difference between the
(degenerate) triplet states and the singlet state. The spin singlet ($
S_{T}=0 $)\ and spin triplet ($S_{T}=1$) states have eigenvalues of $\mathbf{
S}_{T}^{2}$ equal to $0$ and $2$ respectively. Thus, the Hamiltonian (\ref
{spinHamiltonian}) can only have two distinct eigenvalues, and we need to
solve $\lfloor \frac{2}{2}\rfloor +1=2$ equations for $L_0$ and $L_1$. A
convenient $S_{T}=1$ eigenstate is the normalized state $\left\vert \Psi
_{S_{T}=1}^{n=2}\right\rangle \equiv \mathcal{N}\left\vert \Psi (\uparrow
\uparrow )\right\rangle =\mathcal{N}(\left\vert AB\right\rangle \left\vert
\uparrow \uparrow \right\rangle -\left\vert BA\right\rangle \left\vert
\uparrow \uparrow \right\rangle )=\mathcal{N}(\left\vert AB\right\rangle
-\left\vert BA\right\rangle )\left\vert \uparrow \uparrow \right\rangle .$
The normalization constant $\mathcal{N}$ has the value 
\begin{equation*}
\mathcal{N} = (\langle AB|AB\rangle +\left\langle BA\right. \left\vert
BA\right\rangle -\left\langle AB\right. \left\vert BA\right\rangle
-\left\langle BA\right. \left\vert AB\right\rangle )^{-1/2}.
\end{equation*}
Inserting this state into Eq.~(\ref{eq:equal}) yields 
\begin{equation*}
\left\langle \Psi _{S_{T}=1}^{n=2}\right\vert H_{\mathrm{spin}}\left\vert
\Psi _{S_{T}=1}^{n=2}\right\rangle =\left\langle \Psi
_{S_{T}=1}^{n=2}\right\vert H\left\vert \Psi _{S_{T}=1}^{n=2}\right\rangle.
\end{equation*}
The spin Hamiltonian's expectation value is immediately found to be $
L_{0}+2L_{1}$, as can be seen from Eq.~(\ref{eq:Hspin-even}). Expanding out
the spatial Hamiltonian's expectation value gives \begin{widetext}
\begin{equation}
L_{0}+2L_{1}=\frac{\left\langle AB\right\vert H\left\vert AB\right\rangle
+\left\langle BA\right\vert H\left\vert BA\right\rangle -\left\langle
AB\right\vert H\left\vert BA\right\rangle -\left\langle BA\right\vert
H\left\vert AB\right\rangle }{\left\langle AB\right. \left\vert
AB\right\rangle +\left\langle BA\right. \left\vert BA\right\rangle
-\left\langle AB\right. \left\vert BA\right\rangle -\left\langle BA\right.
\left\vert AB\right\rangle }  \label{tripletenergy}
\end{equation}
\end{widetext}
which can be evaluated once a choice of orbital states is specified; we do
this in Section~\ref{values} below. To compare with equation (\ref{eq:J}),
we note that this equation specifies the triplet energy $\epsilon _{t}\equiv
L_{0}+2L_{1}$. A second equation is found from the $S_{T}=0$ state $
\left\vert \Psi _{S_{T}=0}^{n=1}\right\rangle =\mathcal{N}\left( \left\vert
\Psi (\uparrow \downarrow )\right\rangle -\left\vert \Psi (\downarrow
\uparrow )\right\rangle \right) =\mathcal{N}\left( |AB\rangle +|BA\rangle
\right) \left( \left\vert \uparrow \downarrow \right\rangle -\left\vert
\downarrow \uparrow \right\rangle \right) $, which leads to 
\begin{widetext}
\begin{equation}
L_{0}=\frac{\left\langle AB\right\vert H\left\vert AB\right\rangle
+\left\langle BA\right\vert H\left\vert BA\right\rangle +\left\langle
AB\right\vert H\left\vert BA\right\rangle +\left\langle BA\right\vert
H\left\vert AB\right\rangle }{\left\langle AB\right. \left\vert
AB\right\rangle +\left\langle BA\right. \left\vert BA\right\rangle
+\left\langle AB\right. \left\vert BA\right\rangle +\left\langle BA\right.
\left\vert AB\right\rangle },  \label{singletenergy}
\end{equation}
\end{widetext}
giving the singlet energy $\epsilon_{s} = L_{0}$. To exhibit the exchange
coupling explicitly, we rewrite the Hamiltonian as 
\begin{eqnarray}
H_{\mathrm{spin}} &=&L_0 + L_1 \mathbf{S} _{A}^{2} + L_1 \mathbf{S} _{B}^{2}
+2L_1 \mathbf{S}_{A}\cdot \mathbf{S}_{B}  \notag \\
&\equiv& K + J\mathbf{S}_{A}\cdot \mathbf{S}_{B}.  \label{eq:Hspin2}
\end{eqnarray}
where $K = L_{0} + (3/2) L_{1}$ and $J = 2L_{1}$. Expression (\ref{eq:J})
follows when we note that $L_{1} =\frac{1}{2}\left( \epsilon _{t}-\epsilon
_{s}\right)$.

\subsection{Three Electron Case}

\subsubsection{Heitler London Model}

\begin{figure}[tbp]
\includegraphics[height=6.5cm,angle=0]{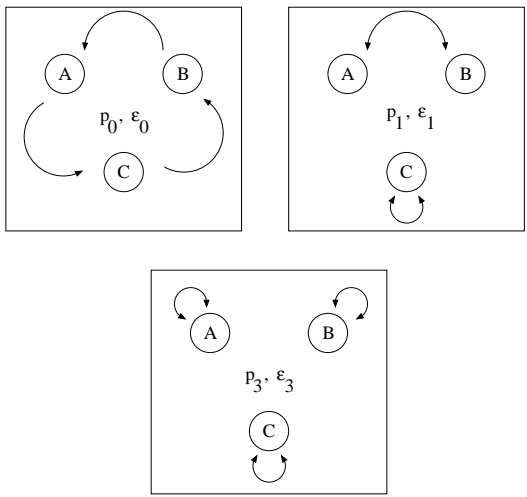}
\caption{Matrix elements relevant to three electron case. Arrows indicate
transition from localized state on initial dot to localized state on final
dot.}
\label{matrixelements3}
\end{figure}

In the three electron case, the possible values that the total spin can take
are $S_{T}=1/2$ (with two, two-dimensional eigenspaces) or $S_{T}=3/2$ (with
a four-dimensional eigenspace). We therefore again need to solve $\lfloor 
\frac{3}{2}\rfloor +1=2$ equations, and it is sufficient to keep only two
constants $L_0 $ and $L_1 $ in $H_{\mathrm{spin}}$, setting $L_2 $ and the
rest to zero. As a convenient state with known $S_{T}=3/2$ we take the
normalized state $|\Psi _{3/2}^{3}\rangle \propto \left\vert \Psi (\uparrow
\uparrow \uparrow )\right\rangle $, so that the energy is $E_{3/2} \equiv
\langle \Psi _{3/2}^{3} \vert H_{\mathrm{spin}}\vert \Psi _{3/2}^{3} \rangle
= L_0 +L_1 (\frac{3}{2})( \frac{5}{2})$. We use $|\Psi _{1/2}^{3}\rangle
\propto \frac{1}{\sqrt{2}} \left( \left\vert \Psi (\uparrow \downarrow
\uparrow )\right\rangle -\left\vert \Psi (\downarrow \uparrow \uparrow
)\right\rangle \right)$ as a normalized state with known $S_{T}=1/2$ , for
which the energy is $E_{1/2} \equiv \langle \Psi _{1/2}^{3}\vert H_{ \mathrm{
spin}}\vert \Psi _{1/2}^{3}\rangle =L_0 +L_1 (\frac{1}{2})(\frac{3}{2 })$.
Then equating expectation values of Hamiltonian (\ref{spatialHamiltonian})
and Hamiltonian (\ref{spinHamiltonian}), i.e., requiring $\left\langle \Psi
_{S_{T}}^{3}\right\vert H_{\mathrm{spin} }\left\vert \Psi
_{S_{T}}^{3}\right\rangle =\left\langle \Psi _{S_{T}}^{3}\right\vert
H\left\vert \Psi _{S_{T}}^{3}\right\rangle $ for each of our states $
S_{T}=1/2$ and $S_{T} = 3/2$ as in Eq.~(\ref{eq:equal}), we can solve for $
L_0$ and $L_1 $. To do so we need to obtain more explicit expressions for $
\left\langle \Psi _{S_{T}}^{3}\right\vert H\left\vert \Psi
_{S_{T}}^{3}\right\rangle $. We assume that $\phi _{A}({\ \mathbf{r}})$, $
\phi _{B}(\mathbf{r})$ , and $\phi _{C}(\mathbf{r})$ are real and satisfy $
\left\langle A|A\right\rangle =\left\langle B|B\right\rangle =\left\langle
C|C\right\rangle $ and $\left\langle A|B\right\rangle =\left\langle
A|C\right\rangle =\left\langle B|C\right\rangle $ (this is consistent with
our original assumption of rotation, inversion, and exchange invariance).
First, let us normalize $|\Psi _{3/2}^{3}\rangle $: 
\begin{eqnarray*}
|\Psi _{3/2}^{3}\rangle &=& \mathcal{N}[|ABC\rangle +|CAB\rangle +|BCA\rangle
\\
&& -|BAC\rangle -|CBA\rangle -|ACB\rangle ]\left\vert \uparrow \uparrow
\uparrow \right\rangle ,
\end{eqnarray*}
where the normalization constant $\mathcal{N}$ is given by 
\begin{equation*}
\mathcal{N}=\frac{1}{\sqrt{6\left( p_{3}+2p_{0}-3p_{1}\right) }}.
\end{equation*}
The quantities $p_{3}$, $p_{1}$, and $p_{0}$ are given by 
\begin{equation*}
p_{3}=\left\langle ABC\right\vert ABC\rangle ,
\end{equation*}
which is an overlap integral when all three electrons retain the same state
in the bra and ket, 
\begin{equation*}
p_{1}=\left\langle BAC\right\vert ABC\rangle =\left\langle CBA\right\vert
ABC\rangle =\left\langle ACB\right\vert ABC\rangle ,
\end{equation*}
which is an overlap integral when one electron has the same state in the bra
and ket, and 
\begin{equation*}
p_{0}=\left\langle CAB\right\vert ABC\rangle =\left\langle BCA\right\vert
ABC\rangle ,
\end{equation*}
which is an overlap integral when zero electrons have the same state in the
bra and ket -- all three electrons change their states. In evaluating the
matrix element $\langle \Psi _{3/2}^{3}\vert H|\Psi _{3/2}^{3}\rangle $ we
use the notation 
\begin{eqnarray*}
\epsilon_{0} &=&\left\langle CAB\right\vert H\left\vert ABC\right\rangle
=\left\langle BCA\right\vert H\left\vert ABC\right\rangle \\
\epsilon_{1} &=&\left\langle BAC\right\vert H\left\vert ABC\right\rangle
=\left\langle CBA\right\vert H\left\vert ABC\right\rangle \\
&=&\left\langle ACB\right\vert H\left\vert ABC\right\rangle \\
\epsilon_{3} &=&\left\langle ABC\right\vert H\left\vert ABC\right\rangle ,
\end{eqnarray*}
where the physical interpretation is that $\epsilon_{k}$ involves $3-k$
electrons exchanging orbitals (Fig.~\ref{matrixelements3}).

Computing the expectation value of $H $ in the state $|\Psi
_{3/2}^{3}\rangle $ then leads to the result: 
\begin{equation}
E_{3/2} = L_0 +\frac{15}{4}L_1 =\frac{\epsilon_{3}+2\epsilon_{0}-3
\epsilon_{1}}{p_{3}+2p_{0}-3p_{1}}  \label{ST3/2}
\end{equation}

For the case $S_{T}=1/2$, using $|\Psi _{1/2}^{3}\rangle $ an analogous
calculation yields: 
\begin{equation}
E_{1/2} = L_0 +\frac{3}{4}L_1 =\frac{\epsilon_{3}-\epsilon_{0}}{p_{3}-p_{0}}.
\label{ST1/2}
\end{equation}
These equations give $L_0$ and $L_1$ in terms of the $p_i$ and $\epsilon_i$.

To compute the usual exchange coupling, it is useful to rewrite $H_{\mathrm{
spin}}$ as 
\begin{eqnarray}
H_{\mathrm{spin}} &=&(L_{0}+L_{1}\sum_{A\leq i\leq C}\mathbf{S}
_{i}^{2})+2L_{1}\sum_{A\leq i<j\leq C}\mathbf{S}_{i}\cdot \mathbf{S}_{j} 
\notag \\
&\equiv &K+J(\mathbf{S}_{A}\cdot \mathbf{S}_{B}+\mathbf{S}_{A}\cdot \mathbf{S
}_{C}+\mathbf{S}_{B}\cdot \mathbf{S}_{C}).  \label{eq:Hspin3}
\end{eqnarray}
where 
\begin{eqnarray}
K &=&L_{0}+\frac{9}{4}L_{1}  \label{eq:K3} \\
J &=&2L_{1}.  \label{eq:J3}
\end{eqnarray}
Solving for the exchange constant $J=2L_{1}$ we find finally: 
\begin{equation*}
J=\frac{2}{3}\left( E_{3/2}-E_{1/2}\right) .
\end{equation*}
A couple of comments are in order concerning this result. First, the
energies $E_{3/2},E_{1/2}$ can be calculated once the orbitals are
specified, as we do in Section~\ref{values} below. We see that, similar to
the two-electron case, the physical interpretation of the exchange constant
is that (up to a multiplicative factor) it is given by the energy difference
between the $S_{T}=3/2$ and $S_{T}=1/2$ states. Second, note from Eqs.~(\ref
{ST3/2}) and (\ref{ST1/2}) that the value of the exchange constant $J$ is
determined in part by the \textquotedblleft
three-electron-exchange\textquotedblright\ terms of the form $
p_{0}=\left\langle CAB\right. \left\vert ABC\right\rangle $ and $\epsilon
_{0}=\left\langle CAB\right\vert H\left\vert ABC\right\rangle $. It is
apparent that such terms involve a cooperative effect between all three
electrons and hence cannot be seen in two-electron calculations. It follows
that \emph{the presence of the third electron quantitatively changes the
exchange coupling between the other two electrons}.

\subsubsection{Hund-M\"{u}lliken Model}

We have have been working within the HL approximation in which there is one
orbital per quantum dot occupied by a single electron. To check its physical
validity, we make three-electron computations within the HM approximation as
well, in which double occupation of quantum dots is permitted. This leads to
a total of $8+12=20$ basis states in the three spin case ($2^{3}=8$ from the
HL basis and $3\times 2\times 2=12$ double-occupation states). In the HL
approximation, the $8$ states divide into a degenerate four-dimensional $
S=3/2$ subspace with energy $E_{3/2}$ and a degenerate four-dimensional $
S=1/2$ subspace with energy $E_{1/2}$. In the HM case, the degenerate
four-dimensional $S=3/2$ subspace is unaffected by the new double-occupation
states which must all have $S=1/2$; the energy of these four $S=3/2$ states
remains $E_{3/2}$. (The $S=3/2$, $S_{z}=3/2$ state has three spin-up
electrons and so the Hamiltonian cannot mix it with any other state. Since
the other $S=3/2$ states are related by a rotation of the arbitrary spin
axis, they must be eigenstates of the Hamiltonian with the same energy.) The 
$12$ double-occupation states enlarge the $S=1/2$ subspace, which becomes $
16 $-dimensional and has a non-trivial spectrum.

In the HM case, the decomposition (\ref{spinsum}) is no longer
meaningful since the basis states do not necessarily have one spin per
quantum dot.  This complicates the computation of the eigenspectrum of
this $16$ -dimensional space. First, we note that the projection
$S_{z}$ (the number of spin-up electrons) is still a good quantum
number since the Hamiltonian (\ref{spatialHamiltonian}) cannot mix two
states with different numbers of spin-up electrons. The
$16$-dimensional subspace therefore splits into two degenerate
$8$-dimensional $S_{z}=\pm 1/2$ subspaces. The $S_{z}=1/2$ subspace
consists of two HL states and six double-occupation states analogous
to (\ref{eq:downupup}):
\begin{widetext}
\begin{eqnarray}
{|\Psi _{1/2}^{3}\rangle } &{\propto }&\frac{1}{\sqrt{2}}\left( \left\vert
\Psi (\uparrow \downarrow \uparrow )\right\rangle -\left\vert \Psi
(\downarrow \uparrow \uparrow )\right\rangle \right) ,{|\Phi
_{1/2}^{3}\rangle \propto \frac{2}{\sqrt{6}}\left\vert \Psi (\uparrow
\uparrow \downarrow )\right\rangle -\frac{1}{\sqrt{6}}\left( \left\vert \Psi
(\uparrow \downarrow \uparrow )\right\rangle +\left\vert \Psi (\downarrow
\uparrow \uparrow )\right\rangle \right) }, \notag \\
{\left\vert \Psi _{AAB}(\uparrow \downarrow \uparrow )\right\rangle } &{
\propto }&a_{A\uparrow }^{\dag }a_{A\downarrow }^{\dag }a_{B\uparrow }^{\dag
}\left\vert \mathrm{vac}\right\rangle ,\left\vert \Psi _{AAC}(\uparrow
\downarrow \uparrow )\right\rangle ,\left\vert \Psi _{BBA}(\uparrow
\downarrow \uparrow )\right\rangle ,\left\vert \Psi _{BBC}(\uparrow
\downarrow \uparrow )\right\rangle ,\left\vert \Psi _{CCA}(\uparrow
\downarrow \uparrow )\right\rangle ,\left\vert \Psi _{CCB}(\uparrow
\downarrow \uparrow )\right\rangle . \label{eq:8states}
\end{eqnarray}
\end{widetext}
One can construct the $8\times 8$ Hamiltonian in this subspace and
diagonalize it. The eigenstates exhibit degeneracies arising from the
symmetry of the Hamiltonian under the exchange of a pair of
dots. Assuming that our dots are all equivalent, there are three
dot-pair exchange operators that commute with the Hamiltonian:
$E_{A,B}$ that exchanges dots $ A,B$, $E_{B,C}$ that exchanges dots
$B,C$, and $E_{C,A}$ that exchanges dots $C,A$. For instance,
$E_{A,B}\left\vert \Psi _{AAC}(\uparrow \downarrow \uparrow
)\right\rangle =\left\vert \Psi _{BBC}(\uparrow \downarrow \uparrow
)\right\rangle $. We can require that the eigenstates of the
Hamiltonian also be eigenstates of $E_{A,B}$ or $E_{B,C}$ or $
E_{C,A}$. Using our 8 states (\ref{eq:8states}), it is possible to
construct $2$ linearly independent states that are
simultaneous eigenstates of all three exchange operators. The two (unnormalized) eigenstates are
\begin{widetext}
\begin{eqnarray*}
&&(\left\vert \Psi _{AAB}(\uparrow \downarrow \uparrow )\right\rangle
+\left\vert \Psi _{AAC}(\uparrow \downarrow \uparrow )\right\rangle
+\left\vert \Psi _{BBA}(\uparrow \downarrow \uparrow )\right\rangle
+\left\vert \Psi _{BBC}(\uparrow \downarrow \uparrow )\right\rangle
+\left\vert \Psi _{CCA}(\uparrow \downarrow \uparrow )\right\rangle
+\left\vert \Psi _{CCB}(\uparrow \downarrow \uparrow )\right\rangle ) \;\; {\rm and}\\
&&(\left\vert \Psi _{AAB}(\uparrow \downarrow \uparrow )\right\rangle
-\left\vert \Psi _{AAC}(\uparrow \downarrow \uparrow )\right\rangle
-\left\vert \Psi _{BBA}(\uparrow \downarrow \uparrow )\right\rangle
+\left\vert \Psi _{BBC}(\uparrow \downarrow \uparrow )\right\rangle
+\left\vert \Psi _{CCA}(\uparrow \downarrow \uparrow )\right\rangle
-\left\vert \Psi _{CCB}(\uparrow \downarrow \uparrow )\right\rangle ),
\end{eqnarray*}
\end{widetext}
with eigenvalue $+1$ and $-1$, respectively. Each such state turns out to be
an eigenstate of the Hamiltonian with its own non-degenerate energy. The
remaining $6$ members of the $8$-dimensional subspace are not simultaneous
eigenstates of all three exchange operators $E_{A,B}$ and $E_{B,C}$ and $
E_{C,A}$. To ensure that we can nevertheless choose the eigenstates of the
Hamiltonian to be simultaneous eigenstates of $E_{A,B}$ or $E_{B,C}$ or $
E_{C,A}$, the energy eigenstates occur in degenerate pairs that can be
superposed as desired to form eigenstates of the exchange operators. When
the parameters of the spatial Hamiltonian (\ref{spatialHamiltonian}) make
double occupation energetically expensive, one of the degenerate pairs will
be low in energy and will consist mainly of the HL states $|\Psi
_{1/2}^{3}\rangle $ and $|\Phi _{1/2}^{3}\rangle $. In this way, the HM
calculation reduces to the HL result plus high energy double-occupation
states, and (\ref{eq:Hspin3}) still describes the low-energy spin dynamics.

\subsubsection{Unequal Coupling}

We emphasize that Eq.~(\ref{eq:Hspin3}) was derived assuming rotation,
inversion, and exchange symmetry. Exchange symmetry, in particular, is
broken whenever there is unequal coupling between dots, and then the
Hamiltonian can involve more constants. This situation is realized when the
dots are not all equidistant, or when they have been shifted electrically,
as in the case of dots defined by electrodes creating confinement
potentials, \cite{Jacak:book} or when there are unequal tunneling barriers
between different dots.\cite{Burkard:99} For instance, in the case of three
unequally coupled dots the Hamiltonian will have the form 
\begin{equation*}
H_{\mathrm{spin}}=K+J_{AB}\mathbf{S}_{A}\cdot \mathbf{S}_{B}+J_{BC}\mathbf{S}
_{2}\cdot \mathbf{S}_{C}+J_{AC}\mathbf{S}_{A}\cdot \mathbf{S}_{C}
\end{equation*}
if we still assume rotation and inversion invariance. (An external magnetic
field, which has been shown to be instrumental in changing the sign of $J$
in the case of two dots,\cite{Burkard:99,Hu:99} could lead to a very
different Hamiltonian. It would break rotation symmetry, introducing
operators into $H_{\mathrm{spin}}$ like $S_{T}^{z}$.) This three electron
Hamiltonian commutes with the $z$-component of the total spin operator $
\mathbf{S}_{T}$, so they can be simultaneously diagonalized. The state $
|\Psi (\downarrow \downarrow \downarrow )\rangle $ has $S_{T}^{z}=-3/2$ and
energy eigenvalue {$K+(J_{AB}+J_{BC}+J_{AC})/4$}. It is found to be
degenerate with {$\frac{1}{\sqrt{3}}(|\Psi (\downarrow \downarrow \uparrow
)\rangle +|\Psi (\downarrow \uparrow \downarrow )\rangle +|\Psi (\uparrow
\downarrow \downarrow )\rangle )$}, a state with $S_{T}^{z}=-1/2$. There are
two remaining $S_{T}^{z}=-1/2$ eigenvectors, which have the (unnormalized)
forms {$(\frac{J_{BC}-J_{AB}+\tilde{J}}{J_{AB}-J_{AC}}|\Psi (\downarrow
\downarrow \uparrow )\rangle +\frac{J_{BC}-J_{AC}+\tilde{J}}{J_{AB}-J_{AC}}
|\Psi (\downarrow \uparrow \downarrow )\rangle +|\Psi (\uparrow \downarrow
\downarrow )\rangle )$}, {$(\frac{J_{AB}-J_{BC}+\tilde{J}}{J_{AC}-J_{AB}}
|\Psi (\downarrow \downarrow \uparrow )\rangle +\frac{J_{AC}-J_{BC}+\tilde{J}
}{J_{AB}-J_{AC}}|\Psi (\downarrow \uparrow \downarrow )\rangle +|\Psi
(\uparrow \downarrow \downarrow )\rangle )$} and have energies $a+\tilde{J}$, $a-\tilde{J}$ respectively, where $a\equiv -3(J_{AB}+J_{BC}+J_{AC})/2$ and 
$\tilde{J}\equiv
(J_{AB}^{2}+J_{BC}^{2}+J_{AC}^{2}-J_{AC}J_{BC}-J_{AB}J_{AC}-J_{AB}J_{BC})^{1/2}. 
$ The remaining four energy eigenvectors, with $S_{T}^{z}=3/2$ and $
S_{T}^{z}=1/2$, can be obtained from these four by inversion. From these
results it is possible to derive equations analogous to (\ref{ST3/2}) - (\ref{eq:J3}) in the case when $J_{AB}$, $J_{BC}$, and $J_{AC}$ are not equal.

\subsection{Four Electron Case}

In the case of four electrons, the effective Hamiltonian again takes the
form (\ref{spinHamiltonian}). Since four electrons can have $S_{T}=0$, $
S_{T}=1$, or $S_{T}=2$, we must keep three constants $L_0 $, $L_1 $, and $
L_2 $ in $H_{\mathrm{spin}}$. It follows immediately that $H_{ \mathrm{spin}
} $ includes terms of the form $L_2 (\mathbf{S}_{A}\cdot \mathbf{S}_{B})( 
\mathbf{S}_{C}\cdot \mathbf{S}_{D})$ and permutations. Unless $L_2 $ happens
to vanish, \emph{the presence of a fourth electron introduces a
qualitatively new 4-body interaction as well as a quantitative change in the
exchange coupling between the other electrons.}

\begin{figure}[tbp]
\includegraphics[height=4cm,angle=0]{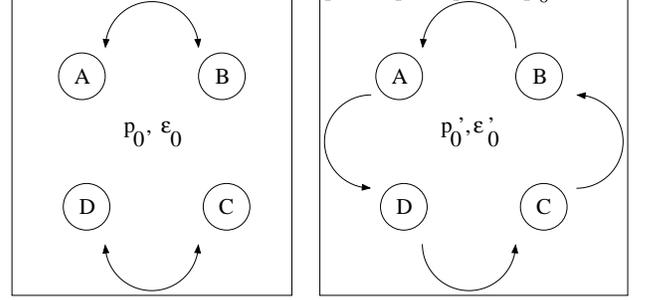}
\caption{Selected matrix elements relevant to four electron case. Arrows
indicate transition from localized state on initial dot to localized state
on final dot.}
\label{matrixelements4}
\end{figure}

We now calculate $L_0 $, $L_1 $, and $L_2 $ just as we calculated $L_0 $ and 
$L_1 $ for three particles. Let us define 
\begin{eqnarray*}
p_{0} &=&\left\langle BADC\right\vert ABCD\rangle \quad
\epsilon_{0}=\left\langle BADC\right\vert H|ABCD\rangle \\
p_{0}^{\prime } &=&\left\langle DABC\right\vert ABCD\rangle \quad
\epsilon_{0}^{\prime }=\left\langle DABC\right\vert H\left\vert
ABCD\right\rangle \\
p_{1} &=&\left\langle ADBC\right\vert ABCD\rangle \quad
\epsilon_{1}=\left\langle ADBC\right\vert H|ABCD\rangle \\
p_{2} &=&\left\langle BACD\right\vert ABCD\rangle \quad
\epsilon_{2}=\left\langle BACD\right\vert H|ABCD\rangle \\
p_{4} &=&\left\langle ABCD\right\vert ABCD\rangle \quad
\epsilon_{4}=\left\langle ABCD|H\right\vert ABCD\rangle
\end{eqnarray*}
where the subscript indicates how many electrons retain the same state in
the bra and the ket, just as in the three electron case. The terms $
\epsilon_{0}$ and $\epsilon_{0}^{\prime }$ involve four-body effects:\ $
\epsilon_{0}$ involves two pairs of electrons exchanging orbitals, and $
\epsilon_{0}^{\prime }$ involves all four electrons exchanging orbitals
cyclically (Fig.~\ref{matrixelements4}).

A convenient state to use for $S_{T}=0$ is $|\Psi^4_0\rangle = \mathcal{\ N}
\left( \left\vert \Psi (\uparrow \downarrow \uparrow \downarrow
)\right\rangle -\left\vert \Psi (\uparrow \downarrow \downarrow \uparrow
)\right\rangle -\left\vert \Psi (\downarrow \uparrow \uparrow \downarrow
)\right\rangle +|\Psi (\downarrow \uparrow \downarrow \uparrow )\rangle
\right) $, keeping in mind the definition (\ref{eq:statedef}).  After normalization, this state yields the singlet energy 
\begin{equation}
E_{0}=L_0 =\frac{\epsilon_{4}-4\epsilon_{1}+3\epsilon_{0}}{
p_{4}-4p_{1}+3p_{0}}.  \label{ST0}
\end{equation}
A convenient state to use for $S_{T}=1$ is $|\Psi^4_1\rangle = \mathcal{N}
\left( \left\vert \Psi (\uparrow \downarrow \uparrow \downarrow
)\right\rangle +\left\vert \Psi (\uparrow \downarrow \downarrow \uparrow
)\right\rangle -\left\vert \Psi (\downarrow \uparrow \uparrow \downarrow
)\right\rangle -|\Psi (\downarrow \uparrow \downarrow \uparrow )\rangle
\right) $. This state, after normalization, yields the triplet energy 
\begin{equation}
E_{1}=L_0 +2L_1 +4L_2 =\frac{\epsilon_{4}-2\epsilon_{2}-\epsilon_{0}+2
\epsilon_{0}^{\prime }}{p_{4}-2p_{2}-p_{0}+2p_{0}^{\prime }}  \label{ST1}
\end{equation}
Finally, a convenient state to use for $S_{T}=2$ is $\left\vert \Psi
(\uparrow \uparrow \uparrow \uparrow )\right\rangle .$ We find for the
quintet energy 
\begin{equation}
E_{2}=L_0 +6L_1 +36L_2 =\frac{\epsilon_{4}-6\epsilon_{2}+8\epsilon_{1}+3
\epsilon_{0}-6\epsilon_{0}^{\prime }}{p_{4}-6p_{2}+8p_{1}+3p_{0}-6p_{0}^{
\prime }}.  \label{ST2}
\end{equation}
Solving, we have 
\begin{eqnarray*}
L_0 &=& E_0 \\
L_1 &=&-\frac{1}{12}\left( E_{2}-9E_{1}+8E_{0}\right) \\
L_2 &=&\frac{1}{24}\left( E_{2}-3E_{1}+2E_{0}\right).
\end{eqnarray*}

We would like to exhibit interaction constants explicitly in the spin
Hamiltonian. We have $\mathbf{S}_{T}=\sum_{i=A}^{D}\mathbf{S}_{i}$, so that 
\begin{equation*}
\mathbf{S}_{T}^{2}=3I+2\sum_{i<j}\mathbf{S}_{i}\cdot \mathbf{S}_{j}
\end{equation*}
while 
\begin{eqnarray*}
(\mathbf{S}_{T}^{2})^{2} &=&(3I+\sum_{A\leq i\neq j\leq D}\mathbf{S}
_{i}\cdot \mathbf{S}_{j})^{2} \\
&=&9I+6\sum_{i\neq j}\mathbf{S}_{i}\cdot \mathbf{S}_{j}+\sum_{i\neq j}
\mathbf{S}_{i}\cdot \mathbf{S}_{j}\sum_{k\neq l}\mathbf{S}_{k}\cdot \mathbf{S
}_{l},
\end{eqnarray*}
and it can be shown that \begin{widetext}
\begin{eqnarray*}
\lefteqn{\sum_{i\neq j}\mathbf{S}_{i}\cdot \mathbf{S}_{j}\sum_{k\neq l}
\mathbf{S}_{k}\cdot \mathbf{S}_{l}} \\
&=&\sum_{i\neq j\neq k\neq l}\left( \mathbf{S}_{i}\cdot \mathbf{S}
_{j}\right) \left( \mathbf{S}_{k}\cdot \mathbf{S}_{l}\right) +4\sum_{i\neq
j\neq l}\left( \mathbf{S}_{i}\cdot \mathbf{S}_{j}\right) \left( \mathbf{S}
_{j}\cdot \mathbf{S}_{l}\right) +2\sum_{i\neq j}\left( \mathbf{S}_{i}\cdot 
\mathbf{S}_{j}\right) \left( \mathbf{S}_{i}\cdot \mathbf{S}_{j}\right)  \\
&=&\sum_{i\neq j\neq k\neq l}\left( \mathbf{S}_{i}\cdot \mathbf{S}
_{j}\right) \left( \mathbf{S}_{k}\cdot \mathbf{S}_{l}\right) +4\left( \frac{1
}{2}\sum_{i\neq j}\mathbf{S}_{i}\cdot \mathbf{S}_{j}\right) +2\left( \frac{9
}{4}-\frac{1}{2}\sum_{i\neq j}\mathbf{S}_{i}\cdot \mathbf{S}_{j}\right) .
\end{eqnarray*}
\end{widetext}
We are led to 
\begin{equation*}
(\mathbf{S}_{T}^{2})^{2}=\frac{27}{2}+7\sum_{i\neq j}\mathbf{S}_{i}\cdot 
\mathbf{S}_{j}+\sum_{i\neq j\neq k\neq l}\left( \mathbf{S}_{i}\cdot \mathbf{S
}_{j}\right) \left( \mathbf{S}_{k}\cdot \mathbf{S}_{l}\right) .
\end{equation*}
The spin Hamiltonian can now be written as 
\begin{eqnarray*}
H_{\mathrm{spin}} &=&K+J\sum_{i<j}\mathbf{S}_{i}\cdot \mathbf{S}
_{j}+J^{\prime }[\left( \mathbf{S}_{A}\cdot \mathbf{S}_{B}\right) \left( 
\mathbf{S}_{C}\cdot \mathbf{S}_{D}\right)  \\
&+&\left( \mathbf{S}_{A}\cdot \mathbf{S}_{C}\right) \left( \mathbf{S}
_{B}\cdot \mathbf{S}_{D}\right) +\left( \mathbf{S}_{A}\cdot \mathbf{S}
_{D}\right) \left( \mathbf{S}_{B}\cdot \mathbf{S}_{C}\right) ].
\end{eqnarray*}
where 
\begin{eqnarray}
K &=&L_{0}+3L_{1}+\frac{27L_{2}}{2}=\frac{-2E_{0}+9E_{1}+5E_{2}}{16}
\label{eq:K4} \\
J &=&2L_{1}+14L_{2}=\frac{-2E_{0}-3E_{1}+10E_{2}}{12}  \label{eq:J4} \\
J^{\prime } &=&8L_{2}=\frac{-2E_{0}-3E_{1}+E_{2}}{3}.  \label{eq:Jprime4}
\end{eqnarray}
Generically, $J^{\prime }$ does not vanish, and four-body interactions
arise. The physical interpretation of the exchange constants as simple
energy differences between different spin multiplets is now lost: we find
energy differences with numerical coefficients that are not intuitively
obvious.

Of central physical importance to us is the relative sizes of the
coefficients $J$ and $J^{\prime}$. This is studied in the next section,
where a HL calculation suggests that $J^{\prime }$ is substantial in
comparison to $J$ in physically important regions of parameter space. We
also find that both coefficients are affected by three ($p_{1}$, $
\epsilon_{1}$) and four-body exchanges ($p_{0}$, $p^{\prime}_{0}$, $\epsilon_{0}$, $\epsilon_{0}^{\prime}$).

In the general case of $2n$ electrons, two-body, four-body,..., $2n$ -body
interaction terms appear in the Hamiltonian. Computing the strengths of the
interactions for larger $n$ is a topic of interest, but we do not address it
here. One expects the strengths of the terms to decrease with the number of
bodies involved.

\section{Model Potential Calculations}

\label{values}

To compute the values of the $L_i$, we select the following specific form
for the one-body potential in (\ref{spatialHamiltonian}): 
\begin{equation}
V(\mathbf{r}) = \frac{1}{2(2l)^6} m \omega _o ^2 |\mathbf{r} - \mathbf{A}|^2
|\mathbf{r} - \mathbf{B}|^2 |\mathbf{r} - \mathbf{C}|^2 |\mathbf{r} - 
\mathbf{D}|^2.  \label{V}
\end{equation}
This potential has a quadratic minimum at each of the vertices of an
equilateral tetrahedron $\mathbf{A} = (0,0,0)$, $\mathbf{B} = (2l\sqrt{\frac{
1}{3}},0,-2l\sqrt{\frac{2}{3}})$, $\mathbf{C} = (-l\sqrt{\frac{1}{3}},l,-2l 
\sqrt{\frac{2}{3}})$, and $\mathbf{D} = (-l\sqrt{\frac{1}{3}},-l,-2l\sqrt{ 
\frac{2}{3}})$. The distance between vertices is $2l$. We select a potential
with four minima so that it can be used in the four electron case without
modification. This facilitates comparison between the two-, three-, and
four-electron cases, and the extra minima do not influence the two- and
three- electron cases in any significant way.

At vertex $\mathbf{A}$, we define the localized Gaussian state 
\begin{equation*}
\phi _A (\mathbf{r}) \equiv \langle \mathbf{r}|A\rangle \equiv \left(\frac{m
\omega _o}{\pi \hbar} \right)^{3/4} \exp \left( -\frac{m \omega _o}{2 \hbar}
|\mathbf{r}-\mathbf{A}|^2 \right)
\end{equation*}
which is the ground state of the quadratic minimum at that vertex. We define
localized states similarly for the other vertices.

The following one-body Hamiltonian matrix elements are needed to evaluate
the coupling constants in $H_{\mathrm{spin}}$ \begin{widetext}
\begin{eqnarray}
\langle A|A\rangle &=&1  \label{onebody} \\
\langle A|h|A\rangle &=& \langle A|\frac{\mathbf{p}^{2}}{2m}+\frac{1}{2}
m\omega _{o}^{2}|\mathbf{r}-\mathbf{A}|^{2}|A \rangle + \langle A|V(\mathbf{
r })-\frac{1}{2}m\omega _{o}^{2}|\mathbf{r}-\mathbf{A}|^{2}|A \rangle  \notag
\\
&=&\hbar \omega _{o}\left[ \frac{3}{2}+\frac{15}{2048}
(63{x_b}^{-3}+280{x_b}^{-2}+320{x_b}^{-1})\right]  \notag \\
\langle A|B\rangle &=&e^{-{x_b}}  \notag \\
\langle A|h|B\rangle &=&\hbar \omega _{o}\left[ \frac{3}{2}e^{-{x_b}} +\frac{1}{
2048} (945{x_b}^{-3}+1680{x_b}^{-2}+936{x_b}^{-1}-1216-880{x_b})e^{-{x_b}}\right] .  \notag
\end{eqnarray}
\end{widetext}
In these equations, we have added and subtracted a harmonic oscillator
potential from the one-body Hamiltonian $h$ for ease of calculation. The
dimensionless tunneling parameter $x_b$ is the square of the ratio of the
interdot distance $2l$ to the characteristic harmonic oscillator width $2
\sqrt{ \hbar/(m \omega_o)}$: 
\begin{equation*}
x_b\equiv m\omega _{o}l^{2}/\hbar,
\end{equation*}
which is also the ratio of the tunneling energy barrier $\frac{1}{2}m
\omega_o^2 l^2$ to the harmonic oscillator ground state energy $\frac{1}{2}
\hbar \omega_o$.

Matrix elements of the Coulomb interaction are given by \begin{widetext}
\begin{eqnarray}
\langle FG|w|UV\rangle &=& \hbar \omega _{o}\left[ 2x_c\sqrt{x_b}\frac{2 \sqrt{2}
}{| \mathbf{f}+\mathbf{u}-\mathbf{g}-\mathbf{v}|}e^{-(1/4)(|\mathbf{f} -
\mathbf{u }|^{2}+|\mathbf{g}-\mathbf{v}|^{2})}{\mathrm{erf}}\left( \frac{| 
\mathbf{f}+ \mathbf{u}-\mathbf{g}-\mathbf{v}|}{2\sqrt{2}}\right) \right] \,
, \,\, |\mathbf{f}+\mathbf{u}-\mathbf{g}-\mathbf{v}|\neq 0  \label{twobody}
\\
\langle FG|w|UV\rangle &=& \hbar \omega _{o}\left[ \frac{4}{\sqrt{\pi }}x_c 
\sqrt{ x_b}e^{-(1/4)(|\mathbf{f}-\mathbf{u}|^{2}+|\mathbf{g}-\mathbf{v} |^{2})}
\right] \, , \,\, |\mathbf{f}+\mathbf{u}-\mathbf{g}-\mathbf{v}|=0.
\label{twobody2}
\end{eqnarray}
\end{widetext}
In these equations, the dimensionless parameter $x_c$ is the ratio of the
Coulomb energy $e^2/(\kappa 2l)$ to the harmonic oscillator ground state
energy $\frac{1}{2} \hbar \omega_o$: 
\begin{equation*}
x_c\equiv e^{2}/(\kappa l\hbar \omega _{o}).
\end{equation*}
The symbols $F$, $G$, $U$, and $V$ take values from the set $\left\{
A,B,C,D\right\} $. The lower case vectors are defined by $\mathbf{f} \equiv 
\sqrt{\frac{m\omega _{o} }{\hbar
}}\mathbf{F}=\frac{\sqrt{x_b}}{l}\mathbf{F} $, etc. The symbol
$\mathrm{erf}(x) = \frac{2}{\sqrt{\pi}} \int_0^x e^{-s^2} ds$ denotes
the error function.

\subsection{Two electrons}

In the case of two electrons, we assume that two of the potential
minima of (\ref{V}) are occupied -- there is an electron at
$\mathbf{A}=(0,0,0)$ and an electron at
${\mathbf{B}}=(2l\sqrt{\frac{1}{3}},0,-2l\sqrt{\frac{2}{3}})$.  In
order to compute $L_{0}$ and $L_{1}$ from Eqs.~(\ref{tripletenergy})
and (\ref{singletenergy}) above, we require only the matrix elements
{$ \left\langle AB\right\vert H\left\vert AB\right\rangle =2\langle
A|h|A\rangle +\langle AB|w|AB\rangle $}, {$\left\langle
AB|AB\right\rangle =1 $}, {$\left\langle AB\right\vert H\left\vert
BA\right\rangle =2\langle A|h|B\rangle \langle A|B\rangle +\langle
AB|w|BA\rangle $}, and {$ \left\langle AB|BA\right\rangle =\langle
A|B\rangle ^{2}$}. (We have simplified using the fact that $\langle
A|h|A\rangle =\langle B|h|B\rangle $ and using the fact that the
wavefunctions are real.) Once $L_{0}$ and $L_{1}$ have been computed,
it is straightforward to obtain $K=L_{0}+(3/2)L_{1}$ and $J=2L_{1}$.

\begin{figure}[tbp]
\includegraphics[height=6.5cm,angle=0]{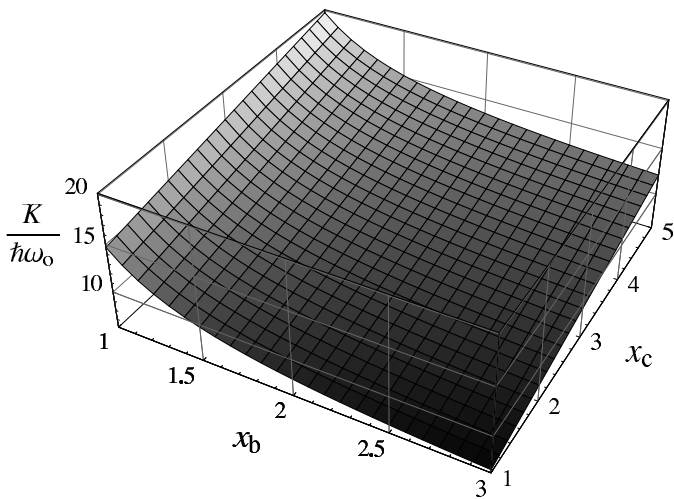} 
\caption{Plot of $K$ as a function of dimensionless tunneling barrier $x_b$
and Coulomb energy $x_c$ in the case of two interacting electrons.}
\label{K2}
\end{figure}

A plot of the energy shift $K$ as a function of $x_b$ (the tunneling energy)
and $x_c$ (the Coulomb energy) is shown in Fig.~\ref{K2} in units of $\hbar
\omega _o$. Following Ref.~\onlinecite{Loss:98}, we estimate realistic
values for $x_b$ and $x_c$ by considering the case of GaAs heterostructure
single dots. An estimated value for $x_b$ is $x_b\equiv
m\omega_{o}^{2}l^{2}/(\hbar \omega_{o}) \approx 1$, since the harmonic
oscillator width $2 \sqrt{\hbar/(m\omega_{o})}$ should be approximately
equal to the distance between dots $2 l$ in a quantum computer. The
parameter $x_c\equiv e^{2}/(\kappa l\hbar \omega _{o}) \approx 1.5$ taking $
\kappa = 13.1$, $\hbar \omega_o = 3\; \mathrm{meV}$ and $x_b \approx 1$. Note
that the energy $K$ increases when the one-electron tunneling barrier energy
decreases and the Coulomb interaction energy increases (i.e. for small $x_b$
and large $x_c$).

\begin{figure}[tbp]
\includegraphics[height=6.5cm,angle=0]{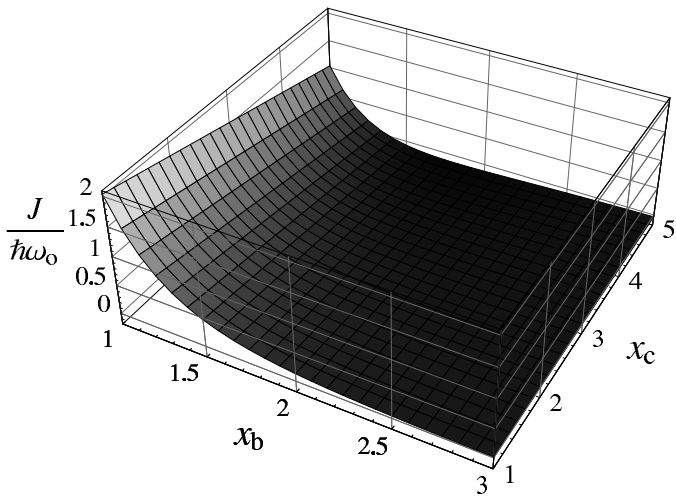}
\caption{Plot of $J$ as a function of $x_b$ and $x_c$ in the case of two
interacting electrons. When $J < 0$ the HL approximation has broken down.}
\label{J2}
\end{figure}

In Fig.~\ref{J2}, we plot the exchange-interaction constant $J$ as a
function of $x_b$ and $x_c$. The plot generally indicates that $J$ increases as
the tunneling barrier decreases ($x_b$ smaller), an intuitively reasonable
result. However, when the Coulomb interaction is particularly strong, as
when $x_c \rightarrow 5$, this trend is violated: $J$ develops a negative
minimum at $x_b\sim 1.5$. The reason is that the $S_T = 3/2$ state has a
totally antisymmetric spatial wavefunction, while the $S_T = 1/2$ state does
not. The antisymmetry tends to reduce the Coulomb repulsion energy between
electrons while increasing the one-electron-tunneling energy. When
parameters are tuned to make the Coulomb repulsion important, the energy of
the $S_T = 3/2$ state dips down, eventually decreasing below the energy of
the $S_T = 1/2$ state. This leads to $J<0$. The negative value of $J$
signals the breakdown of the HL approximation in this region. The exact two
electron ground state is known \cite{Mattis:88} to have $S_{T} = 1/2$, while 
$J <0$ would imply an $S_{T} = 3/2$ ground state. The HL representation of
the $S_{T} = 1/2$ state is simply too rigid to represent the exact ground
state when interactions are strong. The inflexibility of the HL
wavefunctions should be kept in mind when there are strong interactions in
the three electron and four electron case, as well. Fortunately, in our
region of greatest interest, $x_b \approx 1$, $x_c \approx 1.5$, HL results
should be meaningful. Even then, however, it should be kept in mind that the
barrier between minima of the potential (\ref{V}) is shallow, and so the
calculation will become increasingly inaccurate as the minima get close
together. Alternative numerical methods can be found, e.g., in Refs.~\onlinecite{Burkard:99,Burkard:00,Hu:99,Hu:01b}.

\subsection{Three electrons}

\label{threevalues}

In the case of three electrons we assume three of the potential minima in (
\ref{V}) are occupied at, say, $\mathbf{A}=(0,0,0)$, ${\ \mathbf{B}}=(2l 
\sqrt{\frac{1}{3}},0,-2l\sqrt{\frac{2}{3}})$, and $\mathbf{C } =(-l\sqrt{ 
\frac{1}{3}},l,-2l\sqrt{\frac{2}{3}})$. The electrons are therefore arranged
at the corners of an equilateral triangle, and the minimum at ${\mathbf{D}} $
is unoccupied. To solve (\ref{ST3/2}) and (\ref{ST1/2}), we need to evaluate
three-body matrix elements. Details are given in Appendix~\ref{app:3elec}.

\begin{figure}[tbp]
\includegraphics[height=6.5cm,angle=0]{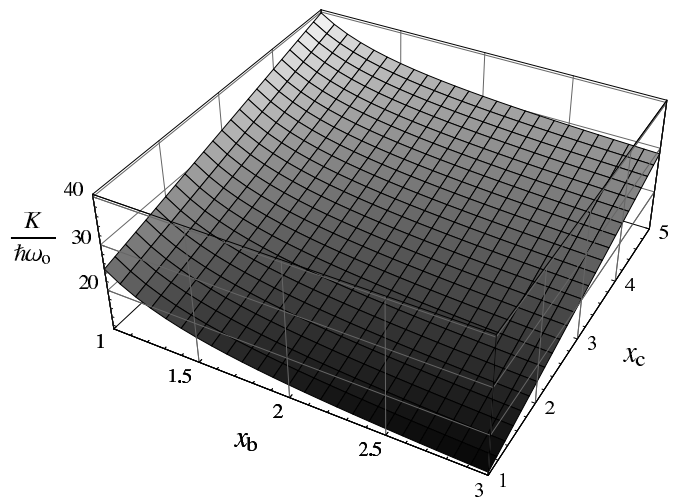}
\caption{Plot of $K$ as a function of $x_b$ and $x_c$ in the case of three
mutually interacting electrons.}
\label{K3}
\end{figure}

\begin{figure}[tbp]
\includegraphics[height=6.5cm,angle=0]{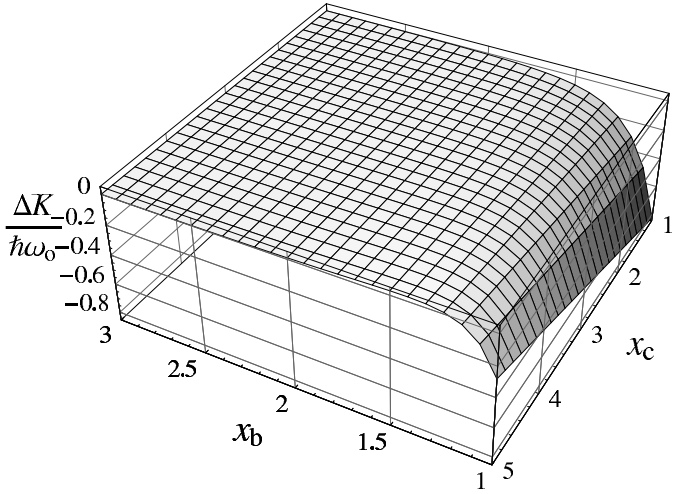}
\caption{Plot of $\Delta K$ as a function of $x_b$ and $x_c$ in the case of
three mutually interacting electrons. Axis directions are reversed from the
preceding figure.}
\label{dK3}
\end{figure}

A plot of the energy shift $K$ as a function of $x_b$ (the tunneling
energy) and $x_c$ (the Coulomb energy) is shown in Fig.~\ref{K3} in
units of $\hbar \omega _o$. The plot's shape is quite similar to that
of Fig.~\ref{K2}. Fig.~\ref{dK3} displays the change $\Delta K$ given by subtracting from $K$ the value that $K$ would take if the three-electron swap matrix elements $\epsilon _{0} = \langle C
A B | H | A B C \rangle $ and $p _0 = \langle C A B | A B C \rangle
$ were zero. The axis directions are reversed in this plot to make its shape
easier to inspect. The figure shows that $\Delta K$ is most important
when one-electron tunneling barrier energy the Coulomb interaction
energy are small in magnitude (small $x_b$ and small $x_c$).

\begin{figure}[tbp]
\includegraphics[height=6.5cm,angle=0]{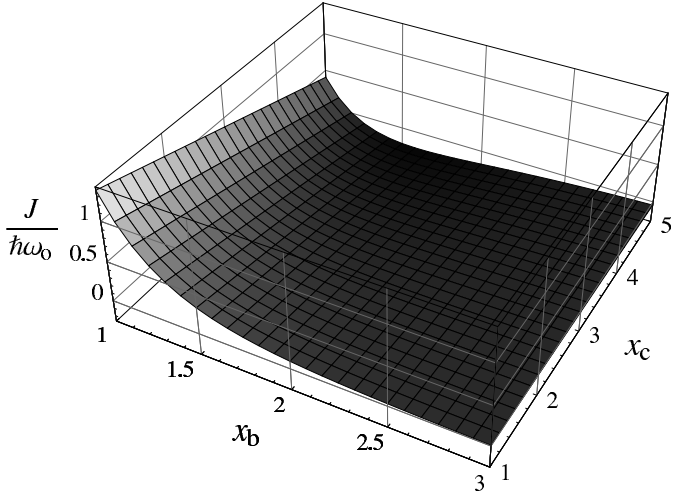}
\caption{Plot of $J$ as a function of $x_b$ and $x_c$ in the case of three
mutually interacting electrons.}
\label{J3}
\end{figure}

In Fig.~\ref{J3}, we plot the exchange-interaction constant $J$ as a
function of $x_b$ and $x_c$. As in the two-body case, the HL
approximation becomes suspect in the regime of strong interactions, so
the region of negative $J$ could just indicate its breakdown. However,
the exact ground state for three or more electrons is not as well
understood as that for two electrons;\cite{Mattis:88} the result $J<0$
cannot be ruled out \textit{a priori}.

\begin{figure}[tbp]
\includegraphics[height=6.5cm,angle=0]{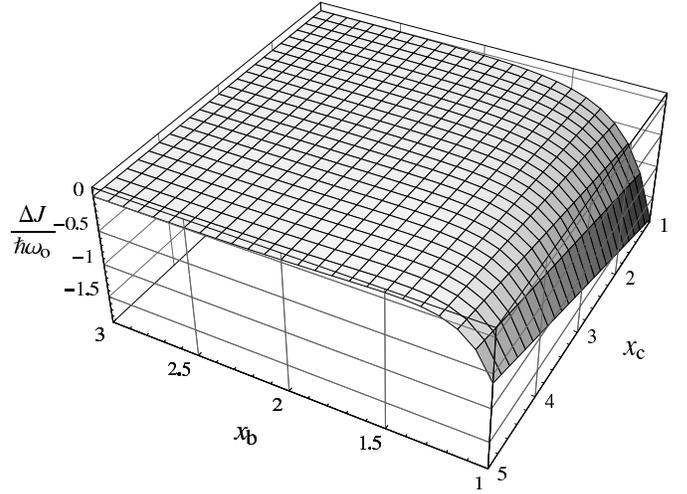}
\caption{Plot of $\Delta J$ as a function of $x_b$ and $x_c$ in the case of
three mutually interacting electrons. Axis directions are reversed from the
preceding figure.}
\label{dJ3}
\end{figure}

Fig.~\ref{dJ3} shows the change $\Delta J$ given by subtracting from $J$ the value that $J$ would take if the three-electron swap matrix elements $\epsilon _0$ and $p_0$ were zero (note that the
axis directions are flipped to make the plot clearer). Comparing the scales
of Figs.~\ref{J3} and \ref{dJ3}, one finds that the three-electron swap
matrix elements can have a powerful influence on $J$.

\begin{figure}[tbp]
\includegraphics[height=6.5cm,angle=0]{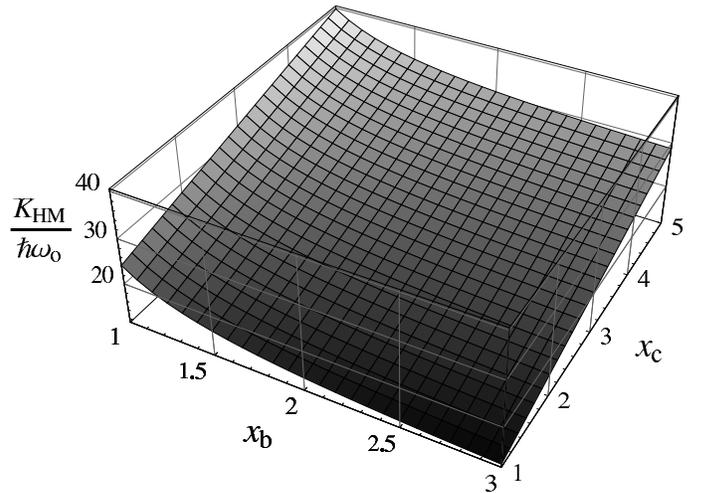}
\caption{Plot of $K$ as a function of $x_b$ and $x_c$ in the case of three
mutually interacting electrons, computed within the HM approximation.}
\label{K3HUNDMULLIKEN}
\end{figure}

\begin{figure}[tbp]
\includegraphics[height=6.5cm,angle=0]{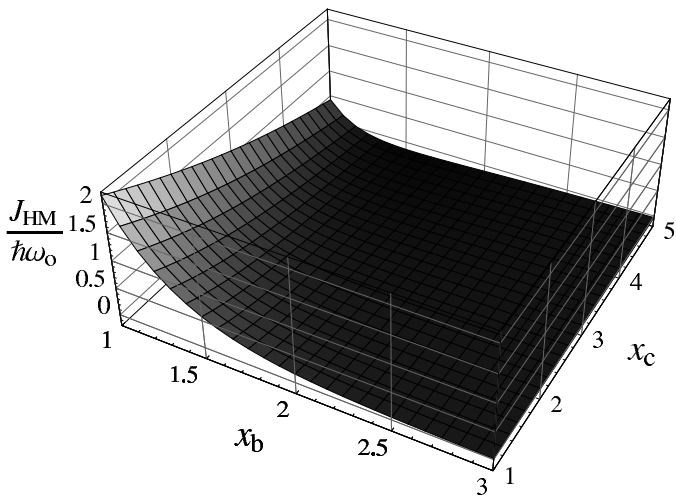}
\caption{Plot of $J$ as a function of $x_b$ and $x_c$ in the case of three
mutually interacting electrons, computed within the HM approximation.}
\label{J3HUNDMULLIKEN}
\end{figure}

To complement our HL results, we have computed the HM spectrum. For
reasonable parameter values ($x_b=1.0,x_c=1.5$), we have found that
the lowest four states of the $16$-dimensional $S=1/2$ subspace are
degenerate and have an energy (that we can call
$E_{1/2,\mathrm{HUND}}$) that is well-separated from that of the
remaining $12$ states with $S=1/2$. These four states are similar in
composition to the four members of the HL $S=1/2$ subspace. The
remaining $12$ states of the HM $S=1/2$ subspace consist mainly of
states with two electrons on a single dot. The four $S=3/2$ states
have an energy that is \emph{in between} $E_{1/2,\mathrm{HUND}}$ and
the energy of the higher-lying $S=1/2$ states. We thus have a
situation that is completely analogous to the one we encountered in
the HL case. It is reasonable to project out the $8$ low energy states
of the HM calculation and compare with the HL
calculation. Figures~\ref{K3HUNDMULLIKEN} and \ref{J3HUNDMULLIKEN}
show the values of $K$ and $J$ for an effective Hamiltonian of the
form (\ref{eq:Hspin3}) that gives this $8$ dimensional low-energy
subspace's spectrum.  These figures should be compared to
Figs.~\ref{K3} and \ref{J3}.

For reasonable parameter values (again, $x_b=1.0,y=1.5$), we find in the HL
approximation that $J=1.9$ for two particles (in units of $\hbar\omega_{o}$), $J=1.2$ for three particles, indicating a change of $-36$\% (or an
absolute change of $-0.7$). In the HM approximation, $J=2.6$ for two
particles and $J=1.8$ for three particles, indicating a change of $-31$\%
(or an absolute change of $-0.8$). Thus, the same effect is seen. The
absolute value of $J$ is larger in the HM case (this is expected since the
basis has increased, leading to a decrease in the ground state energy $
E_{1/2}$ while $E_{3/2}$ stays constant), but the qualitative HL conclusions
are well substantiated.

\subsection{Four Electrons}

\label{fourvalues}

\begin{figure}[tbp]
\includegraphics[height=6.5cm,angle=0]{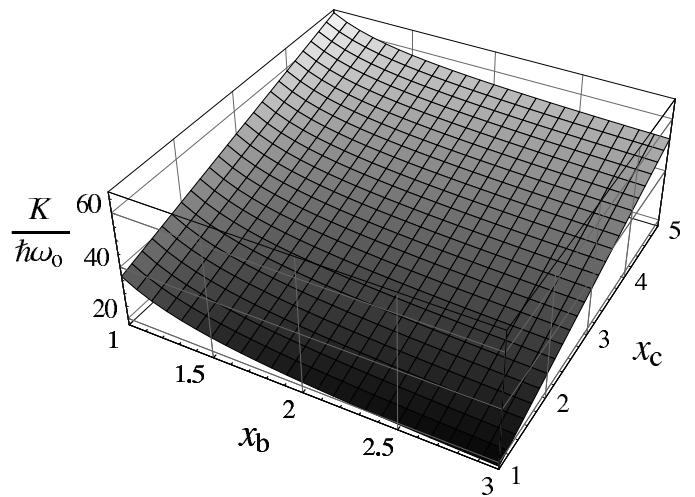}
\caption{Plot of $K$ as a function of $x_b$ and $x_c$ in the case of four
mutually interacting electrons.}
\label{K4}
\end{figure}

\begin{figure}[tbp]
\includegraphics[height=6.5cm,angle=0]{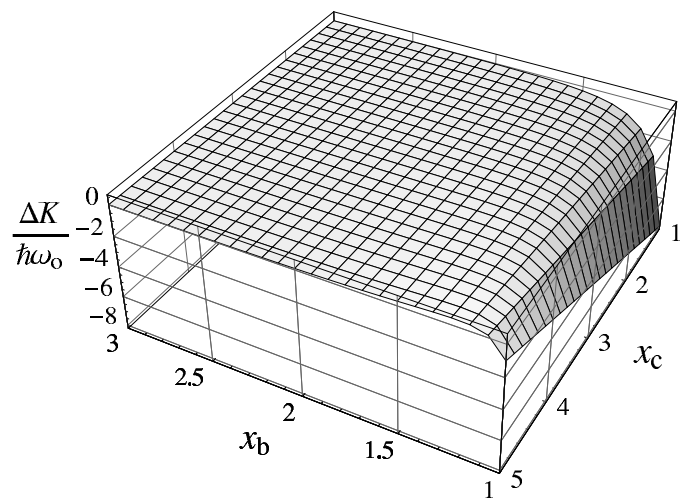}
\caption{Plot of $\Delta K$ as a function of $x_b$ and $x_c$ in the case of four
mutually interacting electrons. Axis directions are reversed from the
preceding figure.}
\label{dK4}
\end{figure}

The actual calculation for four electron case is more involved than that of
the three electron case but identical in procedure. Details are given in
Appendix~\ref{app:4elec}. The resulting quantities $K$ and $\Delta K$ appear
as functions of $x_b$ and $x_c$ in Fig.~\ref{K4} and Fig.~\ref{dK4},
respectively. Here, $\Delta K$ is the value of $K$ minus the value of $K$ obtained by setting to zero both three-body
($p_{1}$, $\epsilon_{1}$) and also four-body ($p_{0}$, $p^{\prime}_{0}$, $
\epsilon_{0}$, $\epsilon_{0}^{\prime}$) matrix elements.

\begin{figure}[tbp]
\includegraphics[height=6.5cm,angle=0]{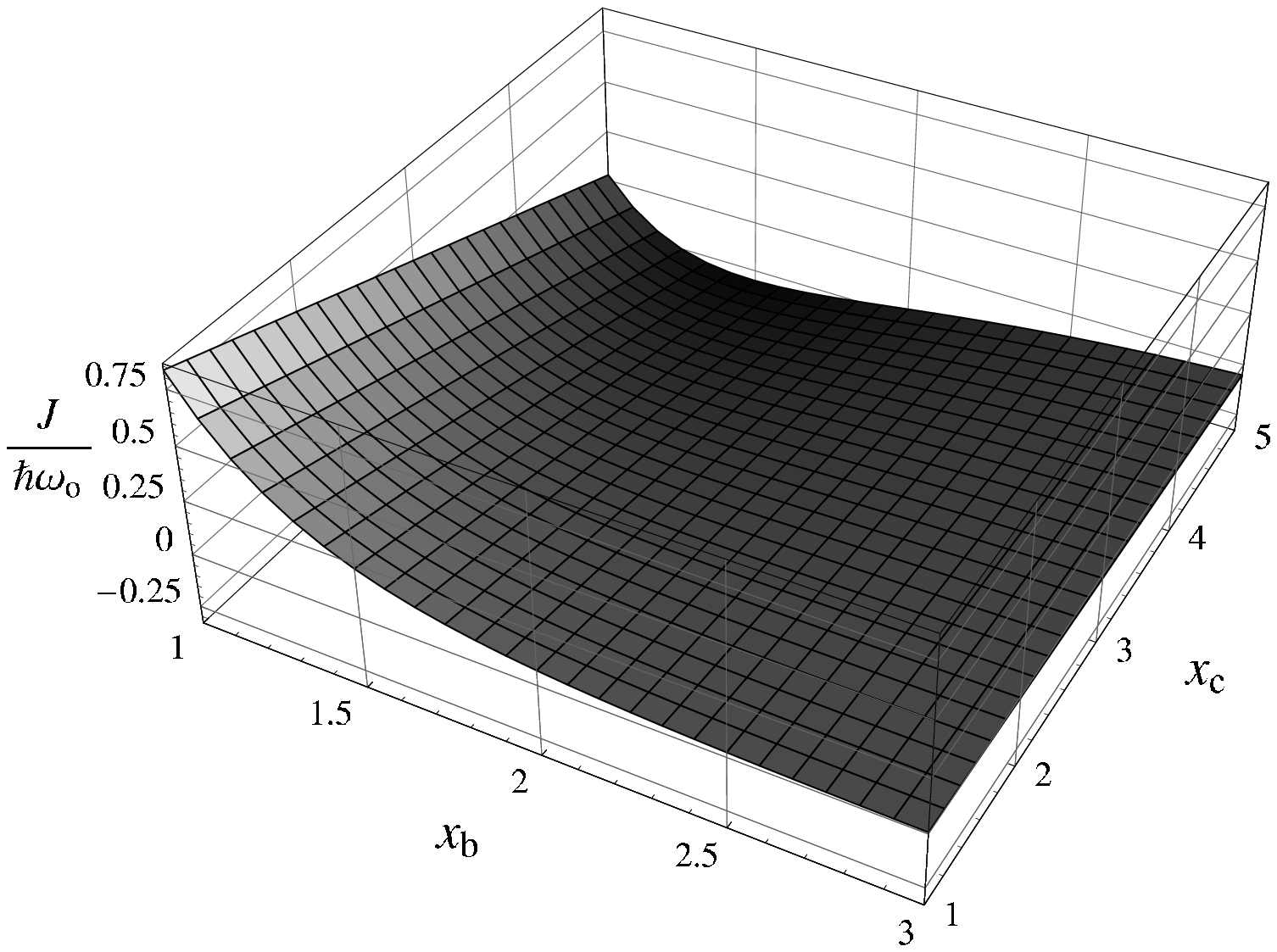}
\caption{Plot of $J$ as a function of $x_b$ and $x_c$ in the case of four
mutually interacting electrons.}
\label{J4}
\end{figure}

\begin{figure}[tbp]
\includegraphics[height=6.5cm,angle=0]{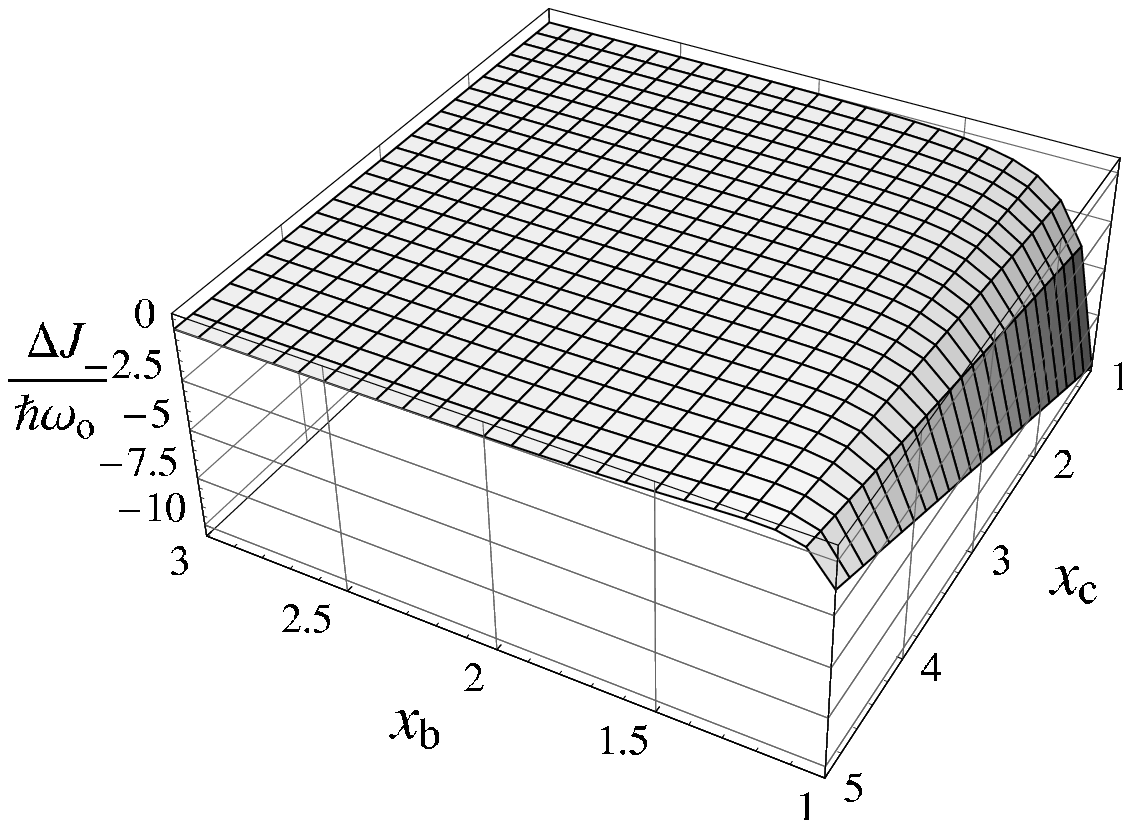}
\caption{Plot of $\Delta J$ as a function of $x_b$ and $x_c$ in the case of four
mutually interacting electrons. Axis directions are reversed from the
preceding figure.}
\label{dJ4}
\end{figure}

\begin{figure}[tbp]
\includegraphics[height=6.5cm,angle=0]{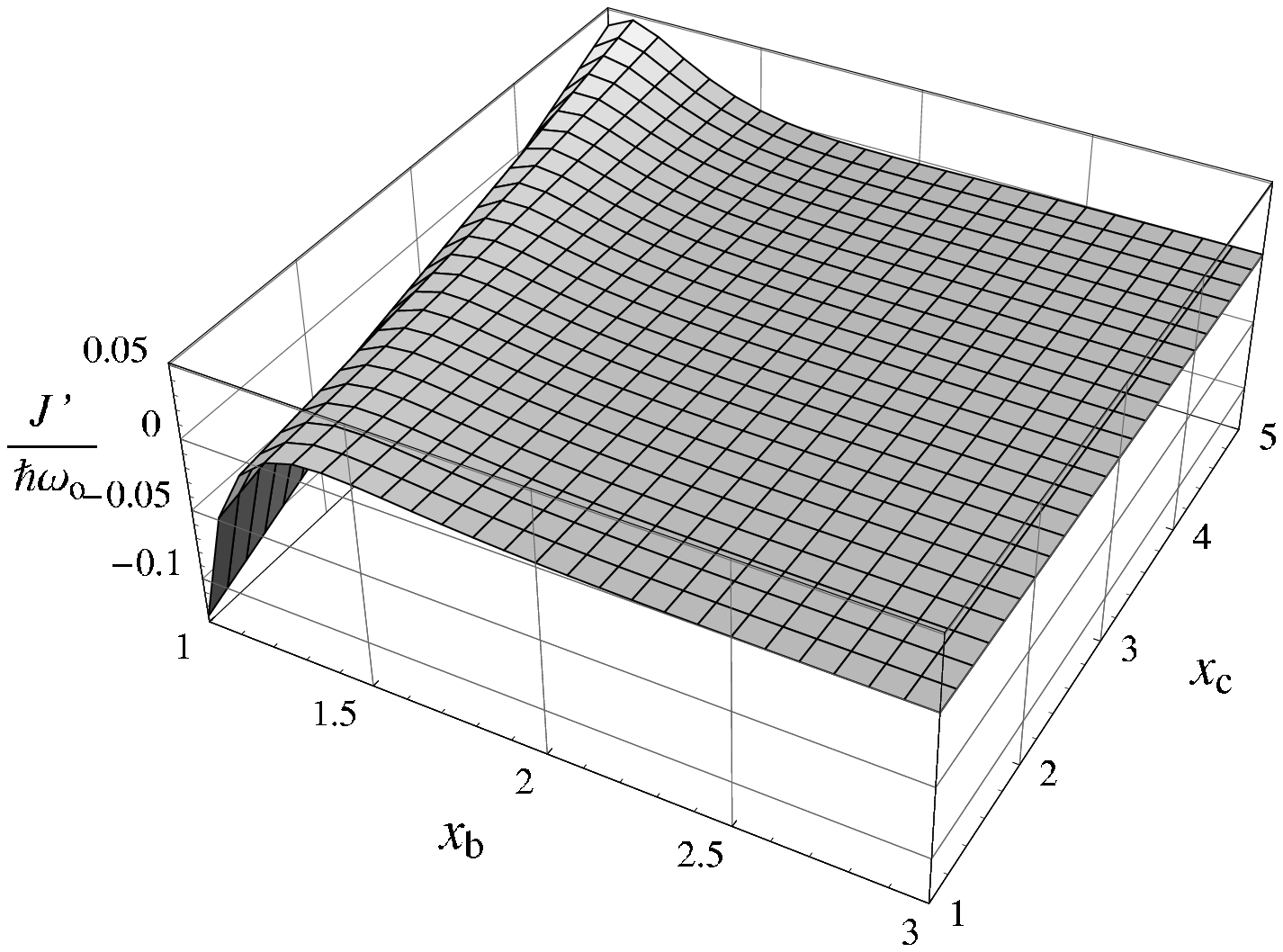}
\caption{Plot of $J^{\prime}$ as a function of $x_b$ and $x_c$ in the case of
four mutually interacting electrons.}
\label{Jp4}
\end{figure}

\begin{figure}[tbp]
\includegraphics[height=6.5cm,angle=0]{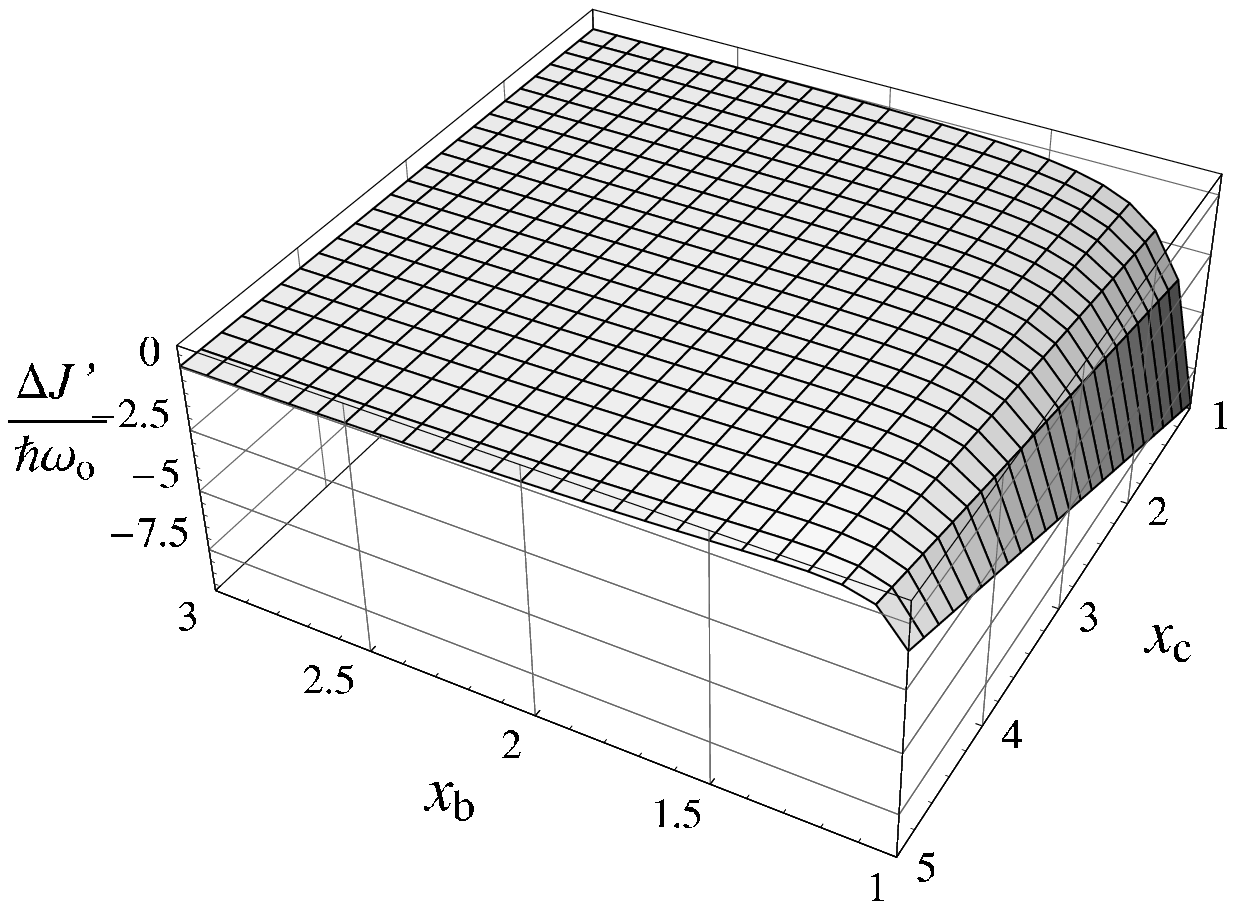}
\caption{Plot of $\Delta J^{\prime}$ as a function of $x_b$ and $x_c$ in the
case of four mutually interacting electrons. Axis directions are reversed from the
preceding figure.}
\label{dJp4}
\end{figure}

The behavior of the exchange-interaction constant $J$ as a function of
$x_b$ and $x_c$ (Fig.~\ref{J4}) is similar to that of the
three-electron case (Fig.~\ref{J3}). The appearance of $\Delta J$
(Fig.~\ref{dJ4}, given by subtracting from $J$ the value that $J$ would take if the three-body and four-body matrix elements were zero) is also reminiscent of $\Delta J$ in the
three-electron case (Fig.~\ref{dJ3}). On the other hand, $J^{\prime}$
(Fig.~\ref{Jp4}) exhibits rich behavior while $\Delta J^{\prime}$
(Fig.~\ref{dJp4}) is qualitatively similar in form to $\Delta J$ from
the three-electron case.

The interaction constant $J^{\prime}$ can be quite significant compared to $
J $, which is remarkable and requires attention in quantum computer design.
In fact, at the point $x_b = 1$, $x_c = 1.5$, our calculation yields $J =
0.76$ and $J^{\prime} = - 0.11$ so $J^{\prime}/J = - 15 \%$, implying
substantial four-body interactions. We caution, though, that these values
were obtained within a HL approximation that will become inaccurate as $x_b$
decreases and the minima of (\ref{V}) get closer together. Our intention is
to highlight the possible significance of the four-body terms. Such terms
have been observed experimentally in $^3$He,\cite{Roger:83} and Cu$_4$O$_4$
square plaquettes in La$_2$CuO$_4$,\cite{Coldea:01} where $J^{\prime}/J$ was
found to be $\sim 27\% $.

\section{Conclusions}

The exchange interaction between localized electrons is a basic
phenomenon of condensed matter physics, with a history that dates back
to Heisenberg's pioneering work.\cite{Heisenberg:28} The details of
its behavior are of great significance to quantum information
processing using quantum dots.  Here we have considered the effects
that arise when three or more electrons, each localized in a low
energy orbital on a quantum dot, are simultaneously coupled. We have
shown that both quantitative and qualitative effects arise, due to
many-body terms, that modify the standard form of the Heisenberg
exchange interaction. Most significantly, in the case of four coupled
electrons, there is a four-body interaction that is added to the
Heisenberg exchange interaction, and our HL calculations suggest that
it could be strong in physically relevant parameter regimes. This
possibility needs to be considered in electron-spin based quantum
computer design, because, on the one hand, of the problems it could
produce when its presence is unwelcome and, on the other hand, because
of its potential uses in novel designs.  In other designs as well, the
possibility should be considered that many-qubit terms could arise in
the effective qubit Hamiltonian.

\begin{acknowledgments}
A.M. acknowledges the support of a Packard Foundation Fellowship for Science
and Engineering. D.A.L. acknowledges support under the DARPA-QuIST program
(managed by AFOSR under agreement No. F49620-01-1-0468), and the Connaught
Fund. We thank Prof. T.A. Kaplan for useful correspondence.
\end{acknowledgments}

\appendix

\section{Perturbative approach to Heisenberg exchange}

\label{app:pert}

Here we summarize the perturbative approach to deriving corrections to
the Heisenberg exchange interaction. See, e.g.,
Refs.~\onlinecite{Takahashi:77,MacDonald:88,March:85vol1} for more
details.

After second quantization of the Coulomb interaction Hamiltonian (\ref
{spatialHamiltonian}) one arrives at the result 
\begin{eqnarray*}
H &=& \sum_{i,s}\varepsilon
_{i}n_{is}+\sum_{i<j}\sum_{s}t_{ij}a_{is}^{\dagger
}a_{js}+U\sum_{i}n_{is}n_{i,-s} \\
&-& \sum_{i<j}\sum_{s,s^{\prime }}J_{ij}a_{is}^{\dagger }a_{is^{\prime
}}a_{js^{\prime }}^{\dagger }a_{js}
\end{eqnarray*}
where $a_{is}^{\dagger }$ creates an electron with spin $s$ in the $i$th
Wannier orbital $\phi (\mathbf{r}-\mathbf{r}_{i})$, $n_{is}=a_{is}^{\dagger
}a_{is}$ is the number operator, 
\begin{equation*}
t_{ij}=\int \phi ^{\ast }(\mathbf{r}-\mathbf{r}_{i})h(\mathbf{r})\phi ( 
\mathbf{r}-\mathbf{r}_{j})d\mathbf{r}
\end{equation*}
is the hopping energy for $i\neq j$, $\varepsilon _{i}\equiv t_{ii}$ is the
energy of the electron in the $i$th orbital, 
\begin{equation*}
U=\frac{1}{2}e^{2}\int \frac{|\phi (\mathbf{r})|^{2}|\phi (\mathbf{r}
^{\prime })|^{2}}{{|\mathbf{r}-\mathbf{r}^{\prime }}|}d\mathbf{r}d\mathbf{r}
^{\prime }
\end{equation*}
is the on-site interaction energy, and 
\begin{eqnarray*}
J_{ij} &=& e^{2}\int \frac{\phi ^{\ast }(\mathbf{r}-\mathbf{r}_{i})\phi (\mathbf{r}-\mathbf{r}_{j})\phi ^{\ast }(\mathbf{r^{\prime }}-\mathbf{r}_{j})\phi ( \mathbf{r}^{\prime }-\mathbf{r}_{i})}{{|\mathbf{r}-\mathbf{r}^{\prime }}|} d \mathbf{r}d\mathbf{r}^{\prime} \\
&\geq& 0
\end{eqnarray*}
is the customary direct exchange integral (ferromagnetic).

One now assumes $U\gg \varepsilon _{i},t_{ij},J_{ij}$ to ensure that all orbitals are singly occupied. One then
evaluates $H$ in the HL basis 
\begin{equation*}
|s_{1},s_{2},\ldots ,s_{n}\rangle =a_{1s_{1}}^{\dagger }a_{2s_{2}}^{\dagger
}\cdots a_{ns_{n}}^{\dagger }\left\vert \mathrm{vac}\right\rangle
\end{equation*}
in which states differ one from the other only in the distribution of spins $
s_{i}$ in the orbitals $i$. The evaluation is facilitated by noting that the
operators 
\begin{equation*}
S_{i}^{\alpha}\equiv \frac{1}{2}\sum_{s,s^{\prime }}a_{is}^{\dagger }\sigma
_{s,s^{\prime }}^{\alpha}a_{is^{\prime }}
\end{equation*}
are the component of spin $1/2$ operators, where $\sigma _{s,s^{\prime
}}^{\alpha}$ are the matrix elements of the Pauli matrices ($\alpha = x,y,z$
). This allows one to rewrite the exchange term as 
\begin{equation*}
\sum_{s,s^{\prime }}J_{ij}a_{is}^{\dagger }a_{is^{\prime }}a_{js^{\prime
}}^{\dagger }a_{js}=2J_{ij}\mathbf{S}_{i}\cdot \mathbf{S}_{j}+\mathrm{const,}
\end{equation*}
which is the familiar Heisenberg exchange Hamiltonian.

The contribution of the hopping term to $H$ can be neglected in the limit $
t_{ij}/U\rightarrow 0$. However, when $t_{ij}/U\ll 1$ but nonvanishing, it
can be shown,\cite{March:85vol1} using standard perturbation theory in
powers of $1/U$, that the effective hopping Hamiltonian in the HL basis
takes the form 
\begin{equation*}
H_{\mathrm{eff}}=-\frac{H_{h}^{2}}{U}+\frac{H_{h}^{3}}{U^{2}}+\frac{
H_{h}^{4} }{U^{3}}+\ldots ,
\end{equation*}
where $H_{h}=\sum_{i<j}\sum_{s}t_{ij}a_{is}^{\dagger }a_{js}$ is the
original hopping Hamiltonian$,$ which vanishes in the HL basis. The first
order correction $-\frac{H_{h}^{2}}{U}$ gives rise to a term of the form $
\frac{1}{U}\sum_{i<j}|t_{ij}|^{2}\mathbf{S}_{i}\cdot \mathbf{S} _{j}+\mathrm{
const}$, which quantitatively modifies (with opposite sign, i.e.,
antiferromagnetically) the Heisenberg Hamiltonian.\cite{Anderson:63a}
However, it is clear that higher order terms can contribute multi-spin terms
of the form we have considered in this paper. It can be shown \cite
{Takahashi:77} that all odd orders vanish, in agreement with our general
symmetry argument of Section~\ref{general-H}. The term $\frac{
H_{h}^{4}}{U^{3}}$ then gives rise to four-spin interactions of the form $( 
\mathbf{S}_{i}\cdot \mathbf{S}_{j})(\mathbf{S}_{k}\cdot \mathbf{S}_{l})$,
proportional to $t_{ij}t_{jk}t_{kl}t_{li}/U^{3}$, with $i<j<l$, $i<k$, $
k\neq j,l$.\cite{Takahashi:77} This can be interpreted diagramatically as a
cycle in which the electrons interchange dots in the order $i\rightarrow
l\rightarrow k\rightarrow j\rightarrow i$. Thus, perturbation theory shows
that when $t^{4}/U^{3}$ is significant, the four-spin interaction cannot be
neglected.

\section{Details of calculations for three electrons}

\label{app:3elec}

The Hamiltonian (\ref{spatialHamiltonian}) contains 3 one-body terms $h$ and
3 Coulomb interaction terms $w$, and the contribution of each term is given
in (\ref{onebody}) - (\ref{twobody2}). These contributions determine the
parameters $p_{3}$, $p_{1}$, $p_{0}$, $\epsilon_{3}$, $\epsilon_{1}$, and $
\epsilon_{0}$ that appear in equations (\ref{ST3/2}) and (\ref{ST1/2}):

\begin{widetext}
\begin{eqnarray*}
\epsilon_{3} & = & \langle A|h|A\rangle + \langle B|h|B\rangle + \langle
C|h|C\rangle + \langle AB|w|AB\rangle + \langle AC|w|AC\rangle + \langle
BC|w|BC\rangle \\
\epsilon _{1} & = & \langle B|h|A\rangle \langle A|B\rangle + \langle
A|h|B\rangle \langle B|A\rangle + \langle C|h|C\rangle \\
& & + \langle BA|w|AB\rangle + \langle BC|w|AC\rangle \langle A|B\rangle +
\langle AC|w|BC\rangle \langle B|A\rangle  \notag \\
\epsilon _{0} & = & \langle C|h|A\rangle \langle A|B\rangle \langle
B|C\rangle + \langle A|h|B\rangle \langle C|A\rangle \langle B|C\rangle +
\langle B|h|C\rangle \langle C|A\rangle \langle A|B\rangle \\
& & + \langle CA|w|AB\rangle \langle B|C\rangle + \langle CB|w|AC\rangle
\langle A|B\rangle + \langle AB|w|BC\rangle \langle C|A\rangle  \notag
\end{eqnarray*}
\end{widetext}
and 
\begin{eqnarray*}
p_{3} & = & 1 \\
p_{1} & = & \langle A|B\rangle \langle B|A\rangle \\
p_{0} & = & \langle C|A\rangle \langle A|B\rangle \langle B|C\rangle
\end{eqnarray*}

We apply the symmetries of an equilateral triangle, $\langle B|A\rangle =
\langle C|A\rangle = \langle B|C\rangle $, $\langle B|h|A\rangle = \langle
C|h|A\rangle = \langle B|h|C\rangle $, $\langle A|h|A\rangle = \langle
B|h|B\rangle = \langle C|h|C\rangle $ to get the matrix elements not
explicitly listed in (\ref{onebody}) - (\ref{twobody2}). All matrix elements
are functions of $\hbar \omega _o$, $x_b$, and $x_c$, so $L_0$ and $L_1$
obtained from (\ref{ST3/2}) and (\ref{ST1/2}), and $K$ and $J$ obtained from
(\ref{eq:K3}) and (\ref{eq:J3}), are functions of the same quantities.

\section{Details of calculations for four electrons}

\label{app:4elec}

Here, there is an electron in a Gaussian orbital at each of the four
potential minima of (\ref{V}). We use the analytical expressions (\ref
{onebody}) - (\ref{twobody2}) to evaluate the many-body matrix elements that
appear in equations (\ref{ST0}), (\ref{ST1}), and (\ref{ST2}). The
Hamiltonian (\ref{spatialHamiltonian}) contains four one-body terms $h$ and
six Coulomb interaction terms $w$. Taking them all into account, we have 
\begin{widetext}
\begin{eqnarray*}
\epsilon _{4} &=&\langle A|h|A\rangle +\langle B|h|B\rangle +\langle
C|h|C\rangle +\langle D|h|D\rangle +\langle AB|w|AB\rangle +\langle
AC|w|AC\rangle  \\
&&+\langle AD|w|AD\rangle +\langle BC|w|BC\rangle +\langle BD|w|BD\rangle
+\langle CD|w|CD\rangle , \\
\epsilon _{2} &=&\langle B|h|A\rangle \langle A|B\rangle +\langle
A|h|B\rangle \langle B|A\rangle +\langle C|h|C\rangle \langle B|A\rangle
\langle A|B\rangle +\langle D|h|D\rangle \langle B|A\rangle \langle
A|B\rangle  \\
&&+\langle BA|w|AB\rangle +\langle BC|w|AC\rangle \langle A|B\rangle
+\langle BD|w|AD\rangle \langle A|B\rangle  \\
&&+\langle AC|w|BC\rangle \langle B|A\rangle +\langle AD|w|BD\rangle \langle
B|A\rangle +\langle CD|w|CD\rangle \langle B|A\rangle \langle A|B\rangle , \\
\epsilon _{1} &=&\langle A|h|A\rangle \langle D|B\rangle \langle B|C\rangle
\langle C|D\rangle +\langle D|h|B\rangle \langle B|C\rangle \langle
C|D\rangle +\langle B|h|C\rangle \langle D|B\rangle \langle C|D\rangle  \\
&&+\langle C|h|D\rangle \langle D|B\rangle \langle B|C\rangle +\langle
AD|w|AB\rangle \langle B|C\rangle \langle C|D\rangle +\langle AB|w|AC\rangle
\langle D|B\rangle \langle C|D\rangle  \\
&&+\langle AC|w|AD\rangle \langle D|B\rangle \langle B|C\rangle +\langle
DB|w|BC\rangle \langle C|D\rangle +\langle DC|w|BD\rangle \langle B|C\rangle 
\\
&&+\langle BC|w|CD\rangle \langle D|B\rangle , \\
\epsilon _{0} &=&\langle B|h|A\rangle \langle A|B\rangle \langle D|C\rangle
\langle C|D\rangle +\langle A|h|B\rangle \langle B|A\rangle \langle
D|C\rangle \langle C|D\rangle  \\
&&+\langle D|h|C\rangle \langle B|A\rangle \langle A|B\rangle \langle
C|D\rangle +\langle C|h|D\rangle \langle B|A\rangle \langle A|B\rangle
\langle D|C\rangle  \\
&&+\langle BA|w|AB\rangle \langle D|C\rangle \langle C|D\rangle +\langle
BD|w|AC\rangle \langle A|B\rangle \langle C|D\rangle +\langle BC|w|AD\rangle
\langle A|B\rangle \langle D|C\rangle  \\
&&+\langle AD|w|BC\rangle \langle B|A\rangle \langle C|D\rangle +\langle
AC|w|BD\rangle \langle B|A\rangle \langle D|C\rangle +\langle DC|w|CD\rangle
\langle B|A\rangle \langle A|B\rangle , \\
\epsilon _{0}^{\prime } &=&\langle D|h|A\rangle \langle A|B\rangle \langle
B|C\rangle \langle C|D\rangle +\langle A|h|B\rangle \langle D|A\rangle
\langle B|C\rangle \langle C|D\rangle  \\
&&+\langle B|h|C\rangle \langle D|A\rangle \langle A|B\rangle \langle
C|D\rangle +\langle C|h|D\rangle \langle D|A\rangle \langle A|B\rangle
\langle B|C\rangle  \\
&&+\langle DA|w|AB\rangle \langle B|C\rangle \langle C|D\rangle +\langle
DB|w|AC\rangle \langle A|B\rangle \langle C|D\rangle +\langle DC|w|AD\rangle
\langle A|B\rangle \langle B|C\rangle  \\
&&+\langle AB|w|BC\rangle \langle D|A\rangle \langle C|D\rangle +\langle
AC|w|BD\rangle \langle D|A\rangle \langle B|C\rangle +\langle BC|w|CD\rangle
\langle D|A\rangle \langle A|B\rangle .
\end{eqnarray*}
The overlap matrix elements are simpler 
\begin{eqnarray*}
p_{4} &=&1 \\
p_{2} &=&\langle B|A\rangle \langle A|B\rangle  \\
p_{1} &=&\langle D|B\rangle \langle B|C\rangle \langle C|D\rangle  \\
p_{0} &=&\langle B|A\rangle \langle A|B\rangle \langle D|C\rangle \langle
C|D\rangle  \\
p_{0}^{\prime } &=&\langle D|A\rangle \langle A|B\rangle \langle B|C\rangle
\langle C|D\rangle .
\end{eqnarray*}
Analytical forms are then available for all of the matrix elements of $H$
and all the overlap matrix elements using expressions (\ref{onebody}) - (\ref
{twobody2}) and using the tetrahedron symmetries

\begin{eqnarray*}
& &\langle B|A\rangle = \langle C|A\rangle = \langle D|A\rangle = \langle
B|C\rangle = \langle B|D\rangle = \langle C|D\rangle \\
& & \langle B|h|A\rangle = \langle C|h|A\rangle = \langle D|h|A\rangle =
\langle B|h|C\rangle = \langle B|h|D\rangle = \langle C|h|D\rangle \\
& & \langle A|h|A\rangle = \langle B|h|B\rangle = \langle C|h|C\rangle =
\langle D|h|D\rangle .
\end{eqnarray*}
\end{widetext}

With all of the matrix elements of $H$ and the overlap matrix elements in
hand, we evaluate $K$, $J$, and $J^{\prime}$ by solving (\ref{ST0}) - (\ref
{ST2}) and (\ref{eq:K4}) - (\ref{eq:Jprime4}).

\end{document}